\documentstyle[psfig]{mn}
\newif\ifAMStwofonts

\ifoldfss
  \ifCUPmtlplainloaded \else
    \NewTextAlphabet{textbfit} {cmbxti10} {}
    \NewTextAlphabet{textbfss} {cmssbx10} {}
    \NewMathAlphabet{mathbfit} {cmbxti10} {} 
    \NewMathAlphabet{mathbfss} {cmssbx10} {} 
  \fi
  \ifAMStwofonts
    \ifCUPmtlplainloaded \else
      \NewSymbolFont{upmath} {eurm10}
      \NewSymbolFont{AMSa} {msam10}
      \NewMathSymbol{\upi}     {0}{upmath}{19}
      \NewMathSymbol{\umu}     {0}{upmath}{16}
      \NewMathSymbol{\upartial}{0}{upmath}{40}
      \NewMathSymbol{\leqslant}{3}{AMSa}{36}
      \NewMathSymbol{\geqslant}{3}{AMSa}{3E}

      \let\leq=\leqslant \let\leq=\leqslant
      \let\geq=\geqslant \let\geq=\geqslant
    \fi
  \fi
\fi 

\ifnfssone
  \newmathalphabet{\mathit}
  \addtoversion{normal}{\mathit}{cmr}{m}{it}
  \addtoversion{bold}{\mathit}{cmr}{bx}{it}
  \newmathalphabet{\mathbfit} 
  \addtoversion{normal}{\mathbfit}{cmr}{bx}{it}
  \addtoversion{bold}{\mathbfit}{cmr}{bx}{it}
  \newmathalphabet{\mathbfss} 
  \addtoversion{normal}{\mathbfss}{cmss}{bx}{n}
  \addtoversion{bold}{\mathbfss}{cmss}{bx}{n}
  \ifAMStwofonts
    \ifCUPmtlplainloaded \else
      \UseAMStwoboldmath
      \makeatletter
      \new@mathgroup\upmath@group
      \define@mathgroup\mv@normal\upmath@group{eur}{m}{n}
      \define@mathgroup\mv@bold\upmath@group{eur}{b}{n}
      \edef\UPM{\hexnumber\upmath@group}
      \new@mathgroup\amsa@group
      \define@mathgroup\mv@normal\amsa@group{msa}{m}{n}
      \define@mathgroup\mv@bold\amsa@group{msa}{m}{n}
      \edef\AMSa{\hexnumber\amsa@group}
      \makeatother
      \mathchardef\upi="0\UPM19
      \mathchardef\umu="0\UPM16
      \mathchardef\upartial="0\UPM40
      \mathchardef\leqslant="3\AMSa36
      \mathchardef\geqslant="3\AMSa3E

      \let\leq=\leqslant \let\leq=\leqslant
      \let\geq=\geqslant \let\geq=\geqslant
    \fi
  \fi
\fi 

\ifnfsstwo
  \DeclareMathAlphabet{\mathbfit}{OT1}{cmr}{bx}{it}
  \SetMathAlphabet\mathbfit{bold}{OT1}{cmr}{bx}{it}
  \DeclareMathAlphabet{\mathbfss}{OT1}{cmss}{bx}{n}
  \SetMathAlphabet\mathbfss{bold}{OT1}{cmss}{bx}{n}
  \ifAMStwofonts
    \ifCUPmtlplainloaded \else
      \DeclareSymbolFont{UPM}{U}{eur}{m}{n}
      \SetSymbolFont{UPM}{bold}{U}{eur}{b}{n}
      \DeclareSymbolFont{AMSa}{U}{msa}{m}{n}
      \DeclareMathSymbol{\upi}{0}{UPM}{"19}
      \DeclareMathSymbol{\umu}{0}{UPM}{"16}
      \DeclareMathSymbol{\upartial}{0}{UPM}{"40}
      \DeclareMathSymbol{\leqslant}{3}{AMSa}{"36}
      \DeclareMathSymbol{\geqslant}{3}{AMSa}{"3E}

      \let\leq=\leqslant \let\leq=\leqslant
      \let\geq=\geqslant \let\geq=\geqslant
    \fi
  \fi
\fi 

\ifCUPmtlplainloaded \else
  \ifAMStwofonts \else 
    \def\upi{\pi}
    \def\umu{\mu}
    \def\upartial{\partial}
  \fi
\fi

\title[High metallicity giant HII regions]{A comprehensive study of reported 
high metallicity giant HII regions. 
I. Detailed abundance analysis}
\author[M. Castellanos et al.]
       {Marcelo~Castellanos,$^1$\thanks{E-mail: marcelo@pollux.ft.uam.es} Angeles I.~D{\'\i}az,$^1$ and
Elena~Terlevich$^2$ 
\thanks{Visiting Fellow, IoA, Cambridge}\\
        $^1$Departamento de F{\'\i}sica Te{\'o}rica, C-XI, Universidad Aut{\'o}noma de Madrid, 28049 Madrid, Spain\\
        $^2$INAOE, Tonantzintla, Apdo. Postal 51, 72000 Puebla, M{\'e}xico\\}
\date{Accepted 
      Received ;
      in original form }

\pagerange{\pageref{firstpage}--\pageref{lastpage}}
\pubyear{2001}

\begin{document}

\maketitle

\label{firstpage}

\begin{abstract}

  We present long-slit observations in the optical and near infrared 
of fourteen H{\sevensize II} regions in the spiral galaxies: NGC~628, NGC~925, NGC~1232
and NGC~1637, all of them reported to have solar or oversolar abundances
according to empirical calibrations. For seven of the observed regions,
ion-weighted temperatures from optical forbidden auroral to nebular line
ratios have been obtained and for six of them, the oxygen abundances derived by
standard methods turn out to be significantly lower than solar.  The other
one, named CDT1 in NGC~1232,  shows an oxygen abundance of 12+log(O/H) = 8.95
$\pm$ 0.20 and constitutes, to the best of our knowledge, the first high
metallicity H{\sevensize II} region for which accurate line temperatures, 
and hence elemental abundances, have been derived. 

For the rest of the regions no line temperature measurements could be
made and the metallicity has been determined by means of both
detailed photoionisation modelling and the sulphur abundance 
parameter S$_{23}$. Only one of these regions shows values of O$_{23}$ and
S$_{23}$ implying a solar or oversolar metallicity. 

According to our analysis, only two of the observed
regions can therefore be considered  as of high metallicity. The two of them
fit the trends  previously  found in other high metallicity H{\sevensize II} regions, i.e.
N/O and S/O  abundance ratios seem to be higher and lower than solar
respectively.

\end{abstract}


\section{Introduction}
The study of nebular abundances has been widely carried out in low 
metallicity H{\sevensize II} regions, since  they show a high excitation spectrum in 
which the temperature sensitive lines, like [O{\sevensize III}]$\lambda$ 4363 {\AA}\, are
clearly  visible and measurable. This 
allows the determination of the gas electron temperature  and 
eventually the derivation of the ionic abundances of the different elements with 
observable emission lines. Some more complex effects 
have to be taken into account, like the existence of different temperatures for 
different lines, the possible presence of temperature fluctuations along the 
line of sight, corrections for unseen ionisation states, etc... which can be
found from observations or from theoretical models, but they are of secondary 
importance unless a very high accuracy is needed, as can be the case for the 
primordial helium abundance determinations.

The analysis of high metallicity H{\sevensize II} regions is far more complicated since, in general, 
their low excitation makes any temperature sensitive line too weak to be
measured. In many cases, the [O{\sevensize III}]$\lambda$ 5007 {\AA}\ line, which is
tipically  one hundred times more intense than the auroral [O{\sevensize III}]$\lambda$
4363 {\AA}\ one, can be barely seen. The reason for this is that the oxygen
optical lines  act as the main coolant for the nebula. A higher oxygen
abundance leads to a  more effective cooling and, as the gas cools down, the
electron temperature gets  lower and the [O{\sevensize III}] optical 
forbidden lines get weaker. This
results in a well known  anticorrelation between electron temperature --and
hence emission line  strengths of the oxygen lines-- and oxygen abundances.
This anticorrelation is the basis of the empirical calibrations used to derive 
abundances in regions in which the electron temperature cannot be 
directly determined. The one most widely used is that 
proposed by Pagel {\it et al.} (1979). It gives the O/H abundance as a
function  of R$_{23}$ = ([O{\sevensize II}]$\lambda$ 3727 + [O{\sevensize III}]$\lambda\lambda$ 4959, 5007)/H$\beta$
which is,  to first order, independent of ionisation parameter. 
The calibration is empirical at the high excitation (low metallicity) end but, 
 at the low
excitation (high metallicity) one it has to rely on theoretical modelling,
since up to now there are no direct determinations of abundances
for high metallicity regions.  

 Single star photoionisation models show that R$_{23}$ depends both on 
ionisation parameter, U,  and on stellar effective temperature, T$^*$, and different 
assumptions about the effects of metallicity on either nebular ionisation
structure or  ionising temperature have been used by different authors to
define a sequence  of models that would eventually allow the calibration of
the upper branch of  the  R$_{23}$ {\it versus} (O/H) relation. From analyses
of H{\sevensize II} region data,   McCall, Rybski \& Shields (1985) concluded that T$^*$
varied with metallicity while the filling factor was constant, while Dopita \&
Evans (1986) concluded just the opposite: that T$^*$ was constant while U
varied with metallicity. These two different assumptions produced calibrations
 of R$_{23}$ yielding abundances that differ by more than a factor of two.

Theoretical stellar evolution models point to a relation between stellar 
metallicity and effective temperature so that, for a given mass, stars of 
higher metallicities show lower effective temperatures. This fact led McGaugh 
(1991) to produce a new R$_{23}$ calibration based on more realistic 
theoretical models in which the ionisation was provided by stellar clusters 
and the effect of metallicity predicted by Maeder (1990) was taken into 
account through the use of the appropriate stellar atmosphere models.
According to his  models, in the high metallicity branch, R$_{23}$ is
relatively insensitive to both T$^*$ and  U and the models converge to a
unique line in the diagram. It should however be taken into account that
McGaugh's star  clusters correspond to zero age stellar populations which
might not be very representative of the ionising populations of H{\sevensize II} regions.
The evolution of  massive stars is fast and metallicity dependent, and the
cluster ionising temperature might not be a monotonically decreasing function
of age due to the appearance of WR stars (Garc{\'\i}a Vargas \& D{\'\i}az 1994;
Garc{\'\i}a-Vargas, Bressan \& D{\'\i}az 1995).
  
Yet, despite their difficulty, the importance of an accurate determination of 
the abundances of high metallicity H{\sevensize II} regions cannot be overestimated since
they constitute most of the H{\sevensize II} regions in early spiral galaxies (Sa to Sbc)
and the inner regions of most late  type ones (Sc to Sd) (D{\'\i}az 1989;
Vila-Costas \& Edmunds 1992) without  which our description of the metallicity
distribution in galaxies cannot be  complete. In particular, the effects of
the choice of different calibrations  on the derivation of abundance gradients
can be very important since any abundance  profile fit will be strongly biased
towards data points at the ends of the  distribution. It should be kept in
mind that abundance gradients are widely  used to constrain chemical evolution
models, histories of star formation over galactic discs or galaxy formation
scenarios.

With the aim of deriving accurate values of abundances in this regime we 
have undertaken the observation of H{\sevensize II} regions for which a relatively high
oxygen abundance (solar or oversolar) has been reported in the literature on
the basis of empirical calibrations.  A first work on this series has
already been published (D{\'\i}az {\it et al.} 2000) in which the analysis of
eight H{\sevensize II} regions in the galaxy NGC~4258 was performed.  Two of the
regions analysed had been observed before and their oxygen abundances were
found to be close to solar on the basis of the empirical R$_{23}$
calibration. Our analysis, based on the measurement of the transauroral [S{\sevensize III}] {$\lambda$} 6312
{\AA} line which, combined with the   near infrared [S{\sevensize III}] {$\lambda\lambda$}
9069, 9532 lines, can provide a sulphur  electron temperature, yielded
oxygen abundances lower than solar by a factor of about two.  Now, we have
selected the H{\sevensize II} regions to be observed from the sample of Van Zee {\it et
al.}  (1998) (hereinafter VZ98) who reported detections
of the transauroral [S{\sevensize III}] {$\lambda$} 6312 {\AA} line for some of them. 

In section 2 we describe the observations; the results are presented in 
section 3, the calculated photoionisation models in section 4 and both are discussed in section 5. 
Finally, section 6 summarizes the main conclusions of this work.

%
%
%
%

\section[]{Observations and data reduction}

The observed H{\sevensize II} regions  have been selected from the sample of VZ98 and all of them present oxygen abundances close to
solar or higher, as deduced from empirical calibrations based on optical
forbidden lines. The selected regions are located in four spiral galaxies:
NGC~628, NGC~925, NGC~1232 and NGC~1637 whose main properties are given in
Table 1.

NGC~628 (M 74) is a late-type giant spiral  classified as Sc(s)I (Sandage \& Tammann, 1981) 
and SA(s)c (de Vaucouleurs {\it et al.}, 1991 (hereinafter RC3)). 
Sharina {\it et al.} (1996) derive a distance of 7.3 Mpc by means of BV photometry
of bright  blue supergiants. Several authors have studied the radial abundance
gradients across this galaxy (McCall, Ribsky \& Shields 1985; Bresolin {\it et al.}
1999;  Belley \& Roy 1992; Ferguson, Gallagher \& Wyse 1998; VZ98) as derived
from empirical calibrations. We have observed four H{\sevensize II} regions in this galaxy,
identified as H13, H3, H4, and H5 in Hodge (1976). These regions are located
at a galactocentric distance  of $\sim$ 225\farcs\ and VZ98 derive for them
values of 12 + log(O/H) around 8.7.

NGC~925 is a late-type barred spiral classified as SBc(s)II-III 
(Sandage \& Tammann 1981) and SAB(s)d (RC3). Son \& Davidge (1998) 
derive a distance of 8.6 Mpc by means of VRI photometry of 
red supergiants. This result is in excellent agreement with that derived from Cepheids 
(Silbermann {\it et al.} 1996). The abundance gradients in this galaxy have
been derived by Martin \& Roy (1994), and Zaritsky {\it et al.} (1994) conclude that 
the oxygen  gradient is flatter than those observed in normal galaxies of the
same morphological type. The new observations of VZ98, however, result in a
steepening of the abundance gradient across this galaxy. We have observed four
of their H{\sevensize II} regions, at a galactocentric distance of $\sim$ 20\farcs\, 
for which VZ98 derive values of 12 + log(O/H) near solar (8.92).

NGC~1232 is a face-on Sc spiral with well defined optical colours 
(RC3) typical of late-type spiral galaxies.  
Van Zee {\it et al.} (1998b) adopt a distance of 21.5 Mpc based on its radial
velocity,  an assumed H$_{0}$ of 75 Km s$^{-1}$ Mpc$^{-1}$ and a Virgocentric
infall model.  Sixteen H{\sevensize II} regions have been analysed by VZ98. We have
observed five of them.

NGC~1637 is a late-type barred spiral classified as SAB(rs)c by 
RC3 and SBc(s)II-III by Sandage \& Tammann (1981). 
Son \& Davidge (1998) derive a distance of 7.8 Mpc. The spiral structure is 
asymmetric with the single outer spiral arm segregated from the inner disc
(Ryder \& Dopita 1993). Fifteen H{\sevensize II} regions in this galaxy have been observed
by VZ98. We have observed one region from their sample which fits our selection
criteria.

Our spectrophotometric observations were obtained with the 4.2m William 
Herschel Telescope at the Roque de los Muchachos Observatory, in 1999 November 8,
using the ISIS double spectrograph, with the EEV12 and TEK4 detectors in 
the blue and red arm respectively. The incoming light was split by the dichroic 
at $\lambda$7500 {\AA}\ .  Two different gratings were used: R300B in the blue
arm and R600R in the red arm, covering  3400 {\AA} in  the blue ($\lambda$3800
to $\lambda$7200) and 800 {\AA} in the near IR ($\lambda$8840 to
$\lambda$9650) and yielding spectral  dispersions of 1.73 {\AA} pixel$^{-1}$
in the blue arm and  0.79 {\AA} pixel$^{-1}$ in the red arm. With a slit 
width of 1\farcs03,  spectral resolutions of {$\sim$}2.0 {\AA} and 1.5 {\AA}
FWHM in the blue and red arms respectively were attained. This is an optimal
configuration which allows the simultaneous observation of a given region in
both frames in a single exposure. Unfortunately, it excludes the [O{\sevensize II}]
$\lambda$3727 line from  observation. We have therefore complemented our data
with those of VZ98 who observed this line on all cases. 

The nominal spatial sampling is 0\farcs4 pixel$^{-1}$ in each frame and the average 
seeing for this night was {$\sim$}1\farcs0. 
A journal of the observations is given in Table 2.

The data were reduced using the IRAF (Image Reduction and Analysis Facility)
package following standard methods. The
two-dimensional wavelength calibration was accurate to 1 {\AA} in all cases by means of 
Cu, Ne and Ar calibration lamps. The two-dimensional frames were flux
calibrated using three spectroscopic  standard stars observed
before and after each programme object with a 3\farcs\ width slit. These standard 
fluxes have been obtained from the most updated version of the 
original Oke's spectra (Oke 1990) and cover the 3200 to 9200 {\AA} range. Standard fluxes
between 9200 and 9650 {\AA} have been obtained from stellar 
atmosphere models. The agreement between the
individual calibration curves was better than 5\% in all cases and a weighted 
mean calibration curve was derived. The spectra were previously 
corrected for atmospheric extinction using a mean extinction curve 
applicable to La Palma observing site.
Regarding background subtraction, the high spectral dispersion used in the
near infrared allowed the almost complete elimination of the night-sky OH
emission lines and, in fact, the observed  $\lambda$9532/$\lambda$9069 ratio
is close to the theoretical value of 2.48 in all cases.
 
Telluric absorptions are negligible in the observed spectra of NGC 628, NGC 1232 and NGC 1637. 
These features have been removed from the spectra 
of the other regions (NGC 925) dividing by a relatively featureless continuum of a subdwarf star 
observed on the same night.
 
\section[]{Results}

Figure 1 shows the spatial distribution of the H$\alpha$ flux along the slit 
for the six different positions observed in the sample, a single one in the case of NGC~628, 
NGC~925 and NGC~1637 and three different ones for NGC~1232. 

Regions H13, H3, H4 and H5 (after Hodge 1976) in NGC~628 are clearly resolved 
at position angle PA=38\degr. 
H3, H4 and H5 are very close together but have been analysed separately due to 
the different excitation conditions evidenced by the [O{\sevensize III}] 
optical forbidden lines.

 Four different regions in NGC~925 are identified at PA=103\degr that we have named 
CDT1, CDT2, CDT3 and CDT4. They show  a high degree of structure. CDT1,
CDT2 and CDT3 are close to the nucleus of the  galaxy. 

Three different slit positions have been observed in NGC~1232. Two regions are 
clearly identified at PA=357\degr that we have named CDT1 
and CDT2. Another region has been observed at PA=78\degr (CDT3) and two regions 
have been resolved at PA=52\degr (CDT4 and CDT5). Region CDT5 in NGC 1232
could not be observed in the near IR frame.

 Finally, one H{\sevensize II} region has been 
observed in NGC~1637 at PA=24\degr (CDT1). All the regions, except CDT5 in 
NGC 1232, had been previously observed by VZ98.

Two representative spectra of our sample are shown in Fig. 2 (region H13 in NGC~628)
and Fig. 3 (region CDT1 in NGC~1232).
Wolf-Rayet features around $\lambda\lambda$ 4680, 5800 {\AA}\ 
are seen in the spectrum of region H13 in NGC~628 
and, to a lesser extent, in those of regions CDT1, CDT3, CDT4 and 
CDT5 in NGC~1232.

\subsection{Line intensities}

Emission line fluxes were measured using the IRAF SPLOT software package, by 
integrating the line intensity over a local fitted continuum. The errors in 
the observed line fluxes have been calculated from the expression
$\sigma_{l}$ = $\sigma_{c}$N$^{1/2}$[1 + EW/(N$\Delta$)]$^{1/2}$, where 
$\sigma_{l}$ is the error in the line flux, $\sigma_{c}$ represents the 
standard deviation in a box near the measured emission line and stands for the 
error in the continuum placement, N is the number of pixels used in the 
measurement of the line flux, EW is the line equivalent width, and
$\Delta$ is the wavelength dispersion in angstroms per pixel. 

The observed line intensities relative to the H$\beta$ line were corrected for 
interstellar reddening according to an average extinction curve (Osterbrock 1989) and 
assuming the Balmer line theoretical values for  case B recombination (Brockehurst 1971).
The presence of an underlying stellar population is clearly evident in the
blue spectra of the observed regions in NGC~925. The H$\gamma$ and H$\delta$
Balmer lines are clearly affected by this stellar absorption. For all the
observed regions an iterative process was applied in order to fit observed and
theoretical Balmer line intensities and to obtain the reddening constant
c(H$\beta$) and its associated error. In all cases, H$\alpha$, H$\gamma$ and
H$\delta$ were fitted within the errors, except for the regions of NGC~925,
for which H$\delta$ seems to be underestimated, and region CDT3 (NGC 1232) for
which both H$\gamma$ and H$\delta$ are clearly underestimated. Reddening
corrected Paschen lines, when measured, are consistent with their theoretical
values. Once the reddening constant has been determined from the Balmer and
Paschen lines,  the errors in the reddening corrected line intensities have
been derived by means of error propagation theory. These line intensities,
together with their corresponding errors,  are given in Table 3 (NGC~628),
Table 4 (NGC~1232), Table 5 (NGC~925) and Table 6 (NGC~1637) for all the
observed H{\sevensize II} regions .  Also given in the tables are the extinction corrected
H$\alpha$ flux,  the H$\beta$  equivalent width and the reddening constant. 
All the regions in our sample were observed by VZ98 and their measured
intensities for the strongest lines are included in our tables for comparison.
In general, both sets of measurements agree 
within the errors.  The major differences arise in  NGC~925 because of the
high degree of structure, that makes difficult an accurate identification of
each region.

\subsection{Physical conditions of the gas} 

Electron densities for each observed region have been derived from the  
[S{\sevensize II}] $\lambda\lambda$ 6717, 6731 {\AA}\ line ratio, following
standard methods (e.g. Osterbrock 1989). They were found to be, in all cases, 
$\leq$ 200 cm$^{-3}$, which corresponds to the low density limit.   

 Different auroral forbidden lines were used for temperature determinations
when possible: [O{\sevensize III}] $\lambda$ 4363 {\AA}, [S{\sevensize III}]
$\lambda$ 6312 {\AA}, [N{\sevensize II}] $\lambda$ 5755 {\AA} and
[S{\sevensize II}] $\lambda\lambda$ 4068,4076 {\AA} together with  their
associated nebular lines. We followed the scheme proposed by Aller (1984) 
using the atomic data by Mendoza \& Zeippen (1983), except in the case of [S{\sevensize
III}] for which the more recent data by Tayal (1997) were used.

For seven H{\sevensize II} regions, 50\% of the sample, it was possible to determine the
[S{\sevensize III}] temperature from the $\lambda$ 6312 {\AA}/$\lambda\lambda$ 9069,9532
{\AA}\ ratio.   These are the four observed regions in NGC~628 and three
of the regions of NGC~1232.  For five of them, at least another line
temperature could be determined and for one of them, H13 in NGC~628,  the four
different temperatures: t(O$^{++}$), t(N$^+$), t(S$^{++}$) and t(S$^{+}$) were
directly derived. 

For region H13 in NGC~628, the four measured temperatures agree within the errors 
yielding a single value of T$_e$ = 9700 K. This is actually expected from
theoretical photoionisation models (Stasi{\'n}ska 1980; Garnett 1992) in this
temperature range. The predicted value of t(O$^{+}$) is, according to the
models, 9900 K. Region H3 in the same galaxy also shows comparable
temperatures of about 10000 K. The rest of the regions for which temperature
determinations are available, show temperatures substantially lower (5400 K
$\leq$ T($S^{++}$)  $\leq$ 8700 K) implying higher abundances.

 In these cases,
photoionisation models predict S$^{++}$  temperatures which are intermediate
between those of O$^{++}$ and O$^{+}$. We have used the linear relation
between t(O$^{++}$) and t(S$^{++}$) (t = 10$^{-4}$ T) found by  Garnett (1992) to
predict the temperature of the O$^{++}$ zone .   For the three regions with
measured temperatures in NGC~1232, predicted  t(O$^{+}$) are fully consistent
with measured t(N$^{+}$). For these regions t(O$^{++}$) was predicted from the
measured t(S$^{++}$) using Garnett's relation and was found to be intermediate
between t(S$^{++}$) and t(O$^+$) with relatively low values (0.45 $\leq$
t($O^{++}$) $\leq$ 0.84).

For the rest of the regions: CDT2 in NGC~1232 and all the observed regions in 
NGC~925 and NGC~1637, it was not possible to obtain a direct measure of the
electron temperature. For these regions an average temperature has been
adopted from the empirical calibration of the sulphur parameter, 
S$_{\rm 23}$ = ([S{\sevensize II}]6717, 6731 + [S{\sevensize III}]9069,
9532)/H$\beta$,  (D{\'\i}az \& P{\'e}rez Montero 2000, (hereinafter DPM00)).
No analysis for region CDT5 in NGC 1232, for which no IR data were obtained,
has been performed.

 Electron densities and temperatures for each of the observed regions are
given in Table 7. Values of the electron temperature derived from measured line
ratios are quoted with their corresponding error. The temperatures derived
from other indirect means are shown without any error assigned.

\subsection{ Chemical abundances}
 
Ionic abundances of the most relevant elements: helium,  oxygen, nitrogen, 
neon  and sulphur have been derived following standard methods (Pagel 
{\it et al.} 1992) and using the corresponding ion-weighted temperatures. 

We have assumed that most of the oxygen is in the first and second ionisation 
stages, and therefore 
O/H = O$^+$/H$^+$ + O$^{++}$/H$^{+}$, and 
N/O= N$^+$/O$^+$. We have also assumed 
S/H = S$^+$/H$^+$ + S$^{++}$/H$^{+}$ which seems to be justified given the relatively low 
estimates of the electron temperature found for most of the observed regions. 

For regions H3, H4 and H5 in NGC~628, no [O{\sevensize II}]{$\lambda$}3727 
{\AA} line fluxes are available since these three regions were integrated together in VZ98. 
Therefore the total abundance of oxygen is derived by means of the empirical
S$_{23}$ calibration.

Mean values of the helium abundance have been determined from the 
He~{\sevensize I} $\lambda$ 4471, 5876 and 6678 {\AA} lines, using the
expressions given by Kunth \& Sargent (1983) and Benjamin, Skillman \& Smits
(1999).  The contribution of neutral helium has been estimated from the
expression: \[ He^{\rm o} + \frac{He^+}{H^+} = \left(1-0.25
\frac{O^+}{O}\right) ^{-1} \frac{He^+}{H^+} \](Kunth \& Sargent 1983).

Ionic and total abundances for all the observed regions are also given in
Table 7. Again, values derived from directly determined temperatures
are quoted with their corresponding error, while those derived from
empirical calibrations are
shown without any error assigned.

\subsection{Wolf-Rayet features}


Relatively prominent Wolf-Rayet features have been observed at 
$\lambda\lambda$ 4660, 5808 {\AA} in region H13 in NGC 628 (see Figure 4).
The observed stellar lines at the $\lambda$ 4660 {\AA} blue bump have been identified as: 
N{\sevensize V} $\lambda\lambda$ 4604, 4620 {\AA}, N{\sevensize III} $\lambda\lambda$ 4634, 4640 {\AA}, a little 
contribution around 4650 {\AA} possibly due to carbon, and the broad He{\sevensize II}
feature at $\lambda$ 4686 {\AA}. Nebular emission lines, [Fe{\sevensize III}]  at $\lambda$
4658 {\AA} and [Ar{\sevensize IV}] at $\lambda$ 4711 {\AA} are also observed. The red bump
of the spectrum shows a broad C{\sevensize IV} feature at $\lambda\lambda$ 5801, 5812
{\AA} while C{\sevensize III} at $\lambda$ 5696 {\AA} is absent. We can  therefore classify
the observed WR stars as WN7 with weak N{\sevensize V} emission, no
presence of C{\sevensize III} at $\lambda$ 5696 {\AA}  and C{\sevensize IV} present but weak. The
observed feature around 4650 {\AA} might belong to C{\sevensize IV}, since we have not
identified any lines from C{\sevensize III} in the spectrum (see Lundstr{\"o}m \& Stenholm
1984). The presence of C{\sevensize IV} in WN spectra is  widely discussed by Conti {\it et al.}
(1983).

%


Another Wolf-Rayet feature has been observed in region CDT3 in NGC 1232 (see Figure 5, top). 
The observed blue bump at $\lambda$ 4660 {\AA} comprises the features of
N{\sevensize III}  $\lambda\lambda$ 4634, 4640~{\AA}, a weak contribution of 
C{\sevensize IV} at
4660 {\AA}  and He{\sevensize II} $\lambda$ 4686 {\AA} lines. The 
N{\sevensize V} lines at
$\lambda\lambda$ 4604, 4620~{\AA} are not detected. Nebular lines,
characteristic of shocked gas, are present, e.g. [Fe{\sevensize III}] at $\lambda$ 4658
{\AA}, [Fe{\sevensize III}] $\lambda$ 5271 {\AA} and [Fe{\sevensize II}] $\lambda$ 5159 {\AA}. Another
evidence for the existence of  shocked gas in this region is the observed
broadening of the nebular  [O{\sevensize I}]{$\lambda\lambda$}6300, 6364 {\AA} forbidden
lines.  The red bump at 5808 {\AA} shows a weak C{\sevensize IV} feature at
$\lambda\lambda$ 5801, 5812 {\AA}. C{\sevensize III} at $\lambda$ 5696 {\AA} is absent.
Other interesting lines typical of WR stars are  found, e.g. weak bumps around
$\lambda$ 5140 {\AA} and 4069 {\AA} due to C{\sevensize II} and C{\sevensize III} and C{\sevensize III} respectively. 

The fact that the combined N{\sevensize III} lines are stronger than the He{\sevensize II} 4686 {\AA} 
line, suggests that the observed WR stars can be classified as WN8. The 
detection of C{\sevensize II} and C{\sevensize III} at 5140{\AA}, could be 
interpreted as a signature of
early or  intermediate WC stars. Hence a mixture of two different WR 
populations could be present in this region.

%


Fainter Wolf-Rayet features are found in regions CDT1, CDT4 and CDT5 in 
NGC 1232. In region CDT1 (see Figure 5, bottom) the N{\sevensize III}$\lambda\lambda$ 4634, 4640 {\AA} 
line strengths indicate that the observed WR stars can be classified 
as WN8. N{\sevensize III}{$\lambda\lambda$}4512, 4528 {\AA} 
are also detected. The red bump 
is not clearly observed.  Wolf-Rayet feature 
intensities and equivalent widths are given in Table 8.

\section{Functional parameters of the observed H{\sevensize II} regions}

Three are the fundamental parameters which control the emission line spectra 
of H{\sevensize II} regions (D\'\i az {\it et al.} 1991):   the shape
of the ionising continuum, the degree of ionisation of the nebula and the
abundance of the gas. This has been parametrised by Z which is scaled with solar, Z$_{\odot}$.
Solar element abundances are adopted as defined in 
Grevesse \& Anders (1989) with refractory elements (Fe, Mg, Al, Ca, Na, Ni) 
depleted by a factor of ten, and Si by a factor of two (Garnett {\it et al.} 1995),
to take into account the depletion onto dust grains. The adopted solar abundance
is then as follows: He/H : -1 ; O/H : -3.08 ; N/H : -3.95 ; S/H : -4.79 ; C/H : -3.44 ;
Ne/H : -3.91 ; Ar/H : -5.44 ; Si/H : -4.75 ; Fe/H : -5.33 ; Mg/H : -5.42 ; Al/H : -6.53 ;
Ca/H : -6.64 ; Na/H : -6.67 ; Ni/H : -6.75, in the notation 12 + log(A/H). 
The ionisation parameter -- {\it i. e.} the ratio of the
ionising photon density to  the particle density -- is a measure of 
the degree of ionisation of the nebula and can be deduced from the ratio of two lines 
of the same element corresponding to two different 
ionisation states,  e.g. [O{\sevensize II}]/[O{\sevensize III}] or 
[S{\sevensize II}]/[S{\sevensize III}]. Alternatively, it can also be 
determined from [O{\sevensize II}/H$\beta$] or [S{\sevensize II}/H$\alpha$] if the 
metallicity of the region is known (D{\'\i}az 1994). 
We have derived $U$ from the expressions given by D{\'\i}az {\it et al.} (2000). In all cases 
the value of U derived from the [O{\sevensize II}]/[O{\sevensize III}] ratio
is systematically  lower than the rest thus implying low effective
temperatures for the ionising  stars. We have
therefore discarded this value  and computed U as the mean of the other three.
These adopted ionisation  parameters -- $logU$ -- are also listed in Table 7 and
their uncertainty is estimated to be around $\pm$ 0.2 dex.

 The shape of the ionising continuum is directly related to the effective temperature 
of the stars that dominate the radiation field responsible for the 
ionisation of the nebula and, in photoionisation models with a single ionising
star, corresponds to the star effective temperature.  A recent version of the
photoionisation code CLOUDY (Ferland 1999) has been used  to estimate the mean
effective temperature of these stars. We have used  Mihalas (1972) NLTE
single-star stellar atmosphere models, with a  plane-parallel geometry and a
constant particle density  through the nebula. The ionising stars are of solar
 metallicity and no effects due to line blanketting or stellar winds are taken into account.

The best-fitting models are obtained by using an optimisation method that
includes the more intense emission line ratios relative to H$\beta$. Input
parameters are the derived oxygen abundance, electron density and ionisation 
parameter given in Table 7. The only varying parameter is the stellar effective
temperature. For the starting model, the relative abundance ratios are taken 
to be solar. Once the best fit is obtained, N/O and S/O abundance ratios are 
varied in order to reconcile, when possible, these ratios with the derived
ones. The computed models can be affected by the assumed refractory elements 
depletion
in the high metallicity regime. In particular Si and, to a lesser extent, 
Fe and Mg can seriously affect the predicted
emission line intensities of [O{\sevensize II}]{$\lambda$} 3727{\AA} and [O{\sevensize III}]{$\lambda$} 5007{\AA}, and
consequently, the derived ion-weighted temperatures (Henry 1993). For example, a change
in the assumed depletion for Si from 2 to 10 relative to solar, implies an enhancement
in the predicted optical forbidden lines of a 50\% in the case of region CDT1 in NGC 1637.  

We have also used CoStar NLTE  single-star stellar atmosphere models (Schaerer
1996) which include  line-blanketing effects and mass loss by stellar winds in
order to compare  the possible differences between the input parameters of
Mihalas and CoStar models at a given metallicity, since  these models are
available only for two metallicities, 0.2Z$_{\odot}$ and
solar. Photoionisation models
for a nebula of a given O/H abundance computed using the two different 
stellar atmosphere sets, which produce the same emission line spectra, 
have values of ionisation parameters and effective temperature which 
 differ by {$\sim$}0.1 dex and 1000 K respectively.

For most regions, consistency is found for effective
temperatures between 34,700 K  and 36,600 K. Predicted line intensities 
and abundances from these single-star photoionisation 
models, as compared to observations, are given in Table 9 for all the regions
in the sample, except for CDT2 and CDT3 in NGC 925, for which no satisfactory 
fit was found.
These regions present serious discrepancies in the observed line intensities 
in comparison to VZ98 data, and therefore have been excluded from our 
subsequent analysis.

\section{Discussion}


\subsection{Observed H{\sevensize II} regions in NGC~628}

The observations of region H13 provide the spectra with the highest signal-to-noise ratio of the 
whole sample. The auroral line of [S{\sevensize III}] at {$\lambda$} 6312 {\AA} was
already detected and measured by VZ98 and the agreement between our measurements and theirs is excellent (as for the rest of the measured lines) as
can be seen from Table 3. We have therefore taken their
observed [O{\sevensize II}] {$\lambda$} 3727 {\AA} line intensity in order to 
perform our
abundance analysis. The four measured electron temperatures (see Table 7)  are
all around 10,000K, except for T(N$^{+}$) which is slightly lower (9,000K $\pm$
700). It must be taken into account, however, that the measurement of the
[N{\sevensize II}]{$\lambda$}5755 {\AA} line could be affected by the presence of the
observed red bump due to Wolf-Rayet stars around 5800 {\AA}. The scheme
proposed by Garnett (1992) seems to describe properly the thermal structure of
this region. This global  isothermal behaviour is perhaps surprising if we
consider that prominent  features due to Wolf Rayet stars are observed.

Indeed, when comparing both the emission line spectra and the derived chemical 
abundances of regions H13 and H3 (see Tables 3 and 7), they are found 
to be nearly identical. In region H13, both standard methods and 
the S$_{23}$ parameter
provide oxygen abundances of 12 + log(O/H) = 8.24 and 8.20 respectively. 
As for region H3, VZ98 do not provide the [O{\sevensize II}]{$\lambda$}3727{\AA} line
intensity since they analyse regions H3, H4 and H5 taken together. Hence, we have calculated the oxygen abundance from the 
S$_{23}$ calibration, which yields a value of 8.23. 
These values are, at least, 0.5 dex lower than previous ones derived 
from empirical calibrations based on the optical forbidden lines 
(VZ98 and references therein). The derived N/O ratios, lower than solar, are consistent 
with the relatively high excitation shown by these regions.
Total and ionic sulphur abundances, calculated for both regions 
from standard methods, are nearly identical and S/O ratios are found to be lower than solar.

A closer look at the observed line intensities reveals one important difference
between H13 and H3;  both the [Ne{\sevensize III}]{$\lambda$}3869{\AA} and 
the [O{\sevensize III}]{$\lambda$}4363{\AA} line intensities are 2 and 1.6 
times
lower respectively in region H13. In fact, when comparing our line intensities
with those from VZ98 for region H13, we clearly see a complete agreement
in all the observed lines except [Ne{\sevensize III}]{$\lambda$}3869{\AA} which is
underestimated in our data by nearly  15\% (0.06 {\it versus} 0.07). This discrepancy 
is a serious one when attempting to proceed further in the analysis of the 
Ne$^{2+}$/O$^{2+}$ ionic ratio. If we choose our observed value, 
we obtain log(Ne$^{2+}$/O$^{2+}$)=-0.96. If, in turn, we adopt the value given
by VZ98, log(Ne$^{2+}$/O$^{2+}$)=-0.89, which seems to be more consistent
with current observations in other galactic and extragalactic H{\sevensize II} regions.
This value is very close to the one derived for region H3 (-0.83).

Regarding the [O{\sevensize III}]{$\lambda$}4363{\AA}, the observed difference in the
line intensity for both regions, H13 and H3, yields a difference of nearly 2,000 K in the 
corresponding mean ion-weighted temperature which translates into a difference of 
0.25 dex in the excitation degree given by the O$^{2+}$/O ratio. Again, we
must be cautious about this result since the measurement of this line in region
H3 offers less realibility due to the lower signal-to-noise ratio. Hence, from
the comparison of both spectra and taking into account the discussed
uncertainties, we can conclude that both regions H13 and H3 show similar 
spectral characteristics and therefore the presence of WR stars in region H13
does not seem to change the global ionisation structure of this region. 

We have calculated NLTE single star photoionisation
models (Mihalas 1972), assuming a plane-parallel geometry and a
constant particle density across the nebula and a gas metallicity, Z,
corresponding to the derived O/H abundance. The best
fitting models reproduce both the observed
emission line spectrum and the O, S and N ionic abundances. The low
estimate of the effective temperature in this region confirms our
conclusions from the comparison of H13 and H3 spectra, that is, the
presence of WR stars in region H13  changes neither the ionisation
structure nor the strength of the ionising radiation for the whole
emitting volume of the region.

Regions H4 and H5 are in contact with each other and show
different excitation conditions from their neighbour region H3.  Their
observed spectra are less excited than that of region H3, because of
the lower photon density that heats these regions. The most interesting
feature in their emission line spectra is the low intensity measured
for the  HeI {$\lambda$} 5876 {\AA} recombination line which indicates
that a large fraction of helium must be in the neutral state, since no
He{\sevensize II} lines are observable at all.  

In these regions, oxygen abundances have been derived by means of the
S$_{23}$ abundance calibration.  Values of 8.31 and 8.34 are found for
regions H4 and H5 respectively.  These values are close to the ones
derived in H3 and H13. Sulphur abundances can be derived by means of
the measured T(S$^{2+}$). Hence, with these measurements, S/O abundance
ratios yield values of -1.56 and -1.67 respectively.  The latter 
corresponds to solar within the errors. The former one, if real, must
be accounted for although it is plausible that, due to the weakness of
the [S{\sevensize III}]{$\lambda$}6312{\AA} line (0.005 relative to H$\beta$), the
measured T(S$^{2+}$) would be underestimated.

Single star photoionisation models point to low ionising temperatures
of around 34,700 K for these two regions.

Prior to our investigation, there was only one H{\sevensize II} region in this
galaxy with the oxygen abundance derived from direct measurements of
the electron temperature: (+292,-020) (VZ98) which lies well below the
oxygen abundance gradient derived by Zaritsky {\it et al.} (1994) from the
calibration of the oxygen optical lines. Corrections to this
calibration on the basis of excitation considerations have been
introduced by Pilyugin (2000, 2001) who suggests that central oxygen
abundances and, therefore,  gradient slopes based on previous,
uncorrected calibrations, could be appreciably overestimated. Figure 6
shows the oxygen radial distribution for NGC~628 as deduced from the
empirical calibrations by Zaritsky {\it et al.}(1994), squares, and Pilyugin
(2001), triangles. Our data are also shown, as filled circles together
with the datum by VZ98. The three regions for which the oxygen
abundances have been derived by the two different calibrations and
direct methods are shown as filled symbols, thus allowing a direct
comparison. Our data and that of VZ98 are slightly below the trend
found by using Pilyugin (2001) calibration which provides a slope
flatter than that found with Zaritsky {\it et al.} (1994) calibration and
providing absolute values for the oxygen abundances lower than theirs
by a factor of about 3. Our data then seem to confirm the suspicion 
that values of abundances for low excitation H{\sevensize II} regions might have been severely overestimated.    
%
%
\subsection{Observed H{\sevensize II} regions in NGC~1232}

Three of the five observed H{\sevensize II} regions in NGC 1232: CDT1, CDT3 and 
CDT4 can be considered as supergiant H{\sevensize II} regions, i. e. have H$\alpha$
luminosities 
greater than 10$^{39}$ erg s$^{-1}$ (Kennicutt 1983). 
Ion weighted temperatures have been derived for the three of them. From our 
analysis,  the adopted trend that
T(A$^{+}$)$>$T(S$^{2+}$) (where A denotes sulphur or nitrogen)  in regions with
an electron temperature lower than 10,000K (Stasi{\'n}ska 1990, Garnett 1992) seems to be confirmed.
Though more reliable observations must be done in order to  confirm this
trend, it seems that a three-zone model nebula can explain  the ionisation
structure of these H{\sevensize II} regions.

\subsubsection{The high metallicity region CDT1}

Region CDT1 deserves special attention because of both its high metal content and the 
presence of Wolf Rayet stars . Derived values for the T(S$^{2+}$) and
T(N$^{+}$) ion-weighted temperatures for these region are 5400K and 6700K with
relative errors of 10\% and  8\% respectively.  This is, to our knowledge, the
first time that electron temperatures are measured in a high metallicity 
extragalactic H{\sevensize II} region. 

A Mihalas model with Z=Z$_{\odot}$, T$_{eff}$=34,900K and 
logU=-2.85 fits adequately the emission line spectra and total
abundance. The ionisation structure of this model is shown in Figure 7
where ionic ratios of O, N and S, together with the average electron
temperature,  are plotted against nebular geometrical depth. It can be seen from this
plot that there is a gradual decrease of the average T$_e$ towards the inner
part of the nebula. Since, at these low temperatures the  O$^{2+}$ emission is
going to take place mostly at IR wavelengths, this  could lead, in principle, to
a certain underestimation of the O$^{2+}$ ratio, and therefore of the
total O abundance. However,  given the low electron temperature of the nebula,
most of the oxygen is in the first ionisation state and hence the
fraction of O$^{2+}$ non accounted for would be very small. 
%

%
 In our analysis, T(O$^{2+}$) has been derived from the observed T(S$^{2+}$) using  Garnett's (1992) 
expression under the assumption of no temperature fluctuations, yielding a value of 
4500 $\pm$ 600 K and hence a derived oxygen abundance of
8.95 $\pm$ 0.20, close to the solar value. Also,
the derived values for log O$_{23}$ and log S$_{23}$ are 0.25 and 0.11
which, according to DPM00, 
implies solar or oversolar abundances. The N/O ratio is found to be
higher than solar by 0.1 dex and, if the above value of O/H  is adopted, S/O is
lower than solar by about the same factor.
However,  the value of the O$^{2+}$/H$^+$
ionic abundance is 12 + log(O$^{2+}$/H) = 8.63, too high in comparison to
what is observed in galactic H{\sevensize II} regions (Shaver {\it et al.} 1983), where the 
excitation degree decreases with increasing metallicity, and somewhat
in disagreement with the ionisation parameter implied by both the
[O{\sevensize II}]/[O{\sevensize III}] and [S{\sevensize II}]/[S{\sevensize III}] line ratios.


A somewhat independent  estimate of the total oxygen abundance can be made 
under the assumption of a solar S/O ratio and a negligible contribution
of S$^{3+}$ to the total S
abundance,  which seems to be justified given the low temperatures
involved (see Fig. 7). In this case values of 
12 + log(O/H) = 8.85 and 12 + log(O$^{2+}$/H) = 8.40 are found, more
in agreement with what is expected from
the observed excitation degree. This implies a value of t(O$^{++}$) =
0.49, slightly higher than that predicted by
Garnett's scheme, but well inside our quoted value. 
Given that  O$^+$/H$^+$ provides a lower limit to the total oxygen
abundance, a conservative value of 12 + log(O/H) = 8.85 $\pm$ 0.20 can be
given.

 This region can therefore 
be considered the first high metallicity H{\sevensize II} region for 
which a reliable total oxygen abundance has been obtained. The fact
that a single star model of T$_{eff}$ $\simeq$ 35,000K adequately reproduces
the emission line spectrum seems to imply that the presence of WR stars
is not appreciably affecting 
the ionisation structure of the region.

\subsubsection{Rest of the regions in NGC 1232}

Three mean ion-weighted temperatures have been measured in the supergiant
H{\sevensize II} region CDT3 (see Table 7). The derived oxygen abundance is 12 + log(O/H)
= 8.56 which is 0.3 dex lower than the derived value from empirical
calibrations (VZ98). The predicted value from the S$_{23}$ abundance 
parameter is 8.51, quite consistent with our derived value from standard
methods. Total N/O and S/O ratios are solar within the errors. The 
Ne$^{2+}$/O$^{2+}$ ionic ratio is 0.11, slightly lower than the mean value derived 
for galactic H{\sevensize II} regions.

This region also shows the presence of WR stars. Again, single star photoionisation 
models, yield a low estimate for the effective temperature of the ionising
radiation (35,000 K) which is combined with a moderately low ionisation
parameter (logU $\simeq$ -2.7).

Similar results are obtained for the supergiant H{\sevensize II} region CDT4. The derived
mean oxygen content range from 8.37, by means of standard methods, to 8.51
by using the S$_{23}$ abundance parameter. The derived N/O and S/O ratios
are solar and oversolar respectively, though this latter one is close to
solar when considering the propagated errors. Single-star photoionisation 
models reproduce satisfactorily the observations with an estimated effective 
temperature of 35,000 K combined with a  moderately low ionisation parameter 
(logU $\simeq$ -2.7).

\subsubsection{Abundance gradient in NGC~1232}

A similar analysis to that made in NGC~628, can be applied to the
observed H{\sevensize II} regions in NGC~1232. Up to now, the oxygen abundance has
been
 derived from direct measurements of the electron temperature for only
 one H{\sevensize II} region: (+135,+114) (VZ98) which lies well below the 
oxygen abundance gradient derived by Zaritsky {\it et al.} (1994) from the
calibration of the oxygen optical lines. 
Figure 8 shows the oxygen radial distribution for NGC~1232 where
symbols have the same meaning 
as in Figure 6. Our data and that of VZ98 follow the same trend found
by using Pilyugin (2001) calibration which provides in this case a
slope similar to that found with Zaritsky {\it et al.} (1994) calibration
but providing absolute values 
for the oxygen abundances lower than theirs by a factor of about 3. A
discrepancy is found, however, for the derived oxygen abundance in the
high metallicity H{\sevensize II} region CDT1, which seems to lie between those
derived according to  both empirical calibrations. Again, our data seem
to confirm the suspicion that values of abundances for low excitation
H{\sevensize II} regions might have been 
severely overestimated.   
%
%
\subsection{Observed H{\sevensize II} regions in NGC~925 and NGC~1637}

No auroral forbidden lines could be measured for the four observed H{\sevensize II} regions in NGC 925,  
due to the low signal-to-noise ratio of their spectra, therefore total
oxygen abundances were derived from the S$_{23}$ parameter. Values of 8.52,
8.71, 8.50  and 8.41 are found for regions CDT1, CDT2, CDT3 and CDT4
respectively. These  derived abundances must be handled with care since, from
the O$_{23}$-S$_{23}$ diagnostic diagram, (DPM00),
these regions  fall near the reversal of the abundance calibration parameter
S$_{23}$.  Another source of uncertainty is the large scatter (20\%) found
between our data and those from VZ98 in the three former regions. These
regions are very  close to the centre of the galaxy, and hence they are
expected to show a  large contribution from an old stellar population. This
fact is  straightforward from the observed spectra as evidenced by the
presence of absorption wings in the Balmer lines and the low values of the  measured
H$\beta$ equivalent widths that range from 6 {\AA} to 47 {\AA}.

Finally, CDT1 in NGC 1637 shows values for log O$_{23}$ and log S$_{23}$ of 0.11 and 
-0.03 respectively. According to these values the metallicity of this region
is likely to be oversolar. The detailed  modelling shows that a Mihalas model
with Z/Z$_{\odot}$=1.80, T$_{eff}$=35,000K  and logU=-3.10 fits adequately the
emission line spectra, except for the near IR [S{\sevensize III}] lines which are overestimated  
by a factor of 2.

\subsection{Global analysis}


The analysis of the derived electron temperatures in four of the
observed GEHR's, indicate that, at least in the cases where the
global emission from the nebula is considered, the mean ion-weighted
temperature of single-ionised atoms (N$^{+}$ and S$^{+}$) is higher than that 
corresponding to twice-ionised species (S$^{2+}$) for electron temperatures 
below 10000 K. Given the high signal-to-noise ratio of these
spectra, and despite the small sample we are dealing with, it can be inferred
that the temperatures in the N$^{+}$ and S$^{+}$ dominated zones vary
as a function of the S$^{2+}$ temperature and, therefore, Garnett's scheme
(Garnett 1992) seems to explain consistently the temperature stratification
in H{\sevensize II} regions.
Another important evidence is the measurement in region
H13 of four ion-weighted temperatures, showing a mean value of 10000 {$\pm$}
700 K, which is again consistent with predictions from single star photoionisation
models (Garnett 1992).

These results are to be compared with the determination of electron temperatures
in other well studied H{\sevensize II} regions. On the whole, the situation is not as clear as
one can infer from our results. The observed H{\sevensize II} regions in NGC 7714 (Gonz{\'a}lez-
Delgado {\it et al.} 1995) and NGC 3310 (Pastoriza {\it et al.} 1993) show a higher excitation 
than the ones in our sample. The temperature stratification
in these regions is reversed in the sense that electron temperatures from 
twice-ionised species are higher than those corresponding to single-ionised atoms.
This behaviour can be understood through the dominance of twice-ionised
species in the cooling of the nebula.
In general, a fair agreement is found between
model predictions and these observations, but several discrepancies are found
between t(S$^{2+}$) and t(O$^{2+}$), specially in NGC 7714. These discrepancies
can be explained in terms of the adopted effective collision strengths for sulphur.
For example, the measured T(S$^{2+}$) in regions B and C are 13,300 K and 12,000 K
respectively, far above T(O$^{2+}$) (11,100 K and 10,100 K respectively). With
the updated collision strengths from Tayal (1997), their derived T(S$^{2+}$) would
decrease to 11,500 K and 10,600 K. These values would be in good agreement, within the
errors, with the ones derived for T(O$^{2+}$).
The same argument can be applied to the Jumbo region in NGC 3310. The measured
T(S$^{2+}$) and T(O$^{2+}$) are 12,500 K and 10,700 K respectively, but T(S$^{2+}$)
would be lowered to 10,900 K if Tayal values are adopted.


%
Regarding global abundances, although 
according to our selection criterion all the observed regions show a high oxygen content (solar or oversolar)
as deduced from empirical calibrations based on the optical oxygen
forbidden lines,  
from this work we can conclude that just two of the observed regions can be
considered as high metallicity H{\sevensize II} regions. One of them, region CDT1 in
NGC 1232, constitutes probably the first high metallicity  extragalactic H{\sevensize II} region for which a direct 
abundance determination has been derived. Kinkel \& Rosa (1994)
also reported a solar oxygen content for region S5 in M101. However, they derived electron
temperatures for low ionisation species ([N{\sevensize II}] and [O{\sevensize II}]) since the lack of proper data on
the [S{\sevensize III}]{$\lambda\lambda$} 9069, 9532{\AA} lines, precluded an accurate derivation of the 
[S{\sevensize III}] temperature. As for the other high metallicity
H{\sevensize II} region, CDT1 in NGC 1637, the oxygen abundance has been derived 
from detailed modelling and therefore, should be considered less reliable.
For the rest of the
regions where an electron temperature has been determined, the abundances
are at least 0.3 dex lower than previously derived. Our values are
in fair agreement with the derived ones from the S$_{23}$ abundance calibration
(DPM00), which seems to be a reliable parameter
for abundance determinations up to metallicities of 0.7Z$_{\odot}$.
Hence, two important questions
arise from our results: the reliability of derived abundances from empirical
calibrations through the oxygen optical forbidden lines and, what seems to be
more important, the reliability of the determination of radial
abundance gradients in external galaxies, which is based on the
previous point.


Figure 9 shows the commonly used diagram of 12+log(O/H) {\em vs}
logR$_{23}$. Crosses correspond to H{\sevensize II} regions in spiral and irregular
galaxies and open circles to H{\sevensize II} galaxies.  Open triangles correspond to high
metallicity H{\sevensize II} regions for which a detailed modelling has been
performed. 
Four different calibrations are shown as labelled in the
plot. The one by Torres-Peimbert {\it et al.} (1989) is fully empirical. That
of Edmunds \& Pagel (1984) is semiempirical and relies heavily on the
model of S5 in M~101 by Shields \& Searle (1978). The other two 
are based on sequences of single star photoionisation models
constructed following two different assumptions: the filling factor
remains constant and the ionising temperature changes with metallicity
(McCall, Rybski \& Shields 1985) or conversely, the ionising
temperature remains constant while the ionisation parameter varies
with metallicity (Dopita \& Evans 1986). Those of our data, 
for which reliable estimates of O/H have been obtained by means 
of direct measurement of the ion-weighted temperatures, are shown 
as filled circles (H13 in NGC 628; CDT1, CDT3 and CDT4 in NGC 1232; 74C, 69C and 5N in
NGC 4258 (D\'\i az et al. 2000)). They all fall below the theoretical calibrations and, in
particular, the position occupied by 
region CDT1 in NGC~1232, with a directly derived oxygen abundance,
points to a calibration
considerably flatter than commonly assumed. All our regions have values
of the excitation parameter P= [O{\sevensize III}]4959+5007/R$_{23}$ (Pilyugin 2001)
less than 0.5 and therefore lie on a low excitation sequence. If this
is the case for a large fraction of giant extragalactic H{\sevensize II} regions,
their oxygen abundances may have been rated too high which 
can have profound implications on the derived gradients. It is clear
that more good quality observations of high metallicity regions are
necessary in order to clarify this important matter.


The new data presented in this work also provides an improved S$_{23}$
calibration. All the observed regions of moderately high metallicity
lie nicely on the calibration of DPM00. The position of region CDT1 in
NGC~1232 seems to indicate that the turnover of the calibration might
be around 12+log(O/H) = 8.7 (Fig. 10). Again, more detailed studies of H{\sevensize II}
regions in this abundance range are needed.

Reliable sulphur abundances have been derived for seven of the observed regions
in the sample. Several authors have discussed a possible gradient in S/O from
Galactic observations (Simpson \& Rubin 1990, Shaver {\it et al.} 1983). In the
case of extragalactic H{\sevensize II} regions, Garnett (1989) conclude, that within the 
observational errors, S/O remains constant as O/H varies. On the other hand, 
D{\'\i}az {\it et al.} (1990)
claim that, despite a global constancy within the errors, differences are
found from galaxy to galaxy (a negative trend in M33 and a positive one
in M101).
In our case, it can be inferred that S/O remains constant as O/H varies with
a dispersion around the solar value of {$\pm$} 0.2 dex (see Fig. 11, left).
Anyway it is interesting
to note that in regions H13 and H3, with relatively low oxygen abundances (0.2 Z$_{\odot}$), 
the S/O abundance ratio seems to be subsolar. The S/O abundance ratio in the high metallicity
H{\sevensize II} region CDT1 in NGC~1232 has been analysed in detail in section 5.2.1.
Clearly, more high spectral resolution observations are needed in order
to infer a possible trend in the S/O abundance ratio.

Fig. 11 (right) shows the N/O {\it versus} O/H abundance diagram for the regions with the most reliable
abundance determinations: H13 in NGC 628, CDT1, CDT3 and CDT4 in NGC 1232. We have also included
regions 74C and 69C in NGC 4258 for which similar data were previously analysed (D{\'\i}az {\it et al.} 
2000). Our values are very close to those obtained by Garnett \& Shields (1987) for M81 H{\sevensize II} regions
with similar oxygen content. Region H13 in NGC 628 shows an undersolar value (-1.08 {\it versus} -0.87)
although larger than the LMC by 0.4 dex (Garnett 2000), which shows a similar oxygen content. 
On the extreme, region CDT1 in
NGC 1232 shows a slightly oversolar value very close to those found for the H{\sevensize II} regions in M51.
In general, our data follow the global trend of increasing N/O with O/H found in HII regions over
galactic discs. The relation between N/O and O/H is difficult to explain once the closed box model
for chemical evolution is abandoned.
Although in principle, abundance ratios between different elements are a good signature of stellar
nucleosynthesis processes, the O/H abundance can be affected by other processes like infall and/or
radial outflows of gas. Also, some amount of oxygen may be locked into dust grains thus disappearing
from the gas phase, which complicates the interpretation of the observed N/O in terms of stellar 
yields.
Anyway the relatively large value of N/O found for the high metallicity region CDT1 in NGC 1232
seems to imply a certain contribution by secondary nitrogen production.

\subsection{Possible evolutionary effects in GEHR's.}

Masegosa (1988) found, from a wide sample
of Giant Extragalactic H{\sevensize II} Regions (GEHR) and H{\sevensize II} galaxies, an increase 
in the 
Ne$^{2+}$/O$^{2+}$ ionic ratio with decreasing H$\beta$ equivalent width, 
i.e., as the ionising cluster turns out to be more evolved,  
though there is a non-negligible dispersion in the observed values. This 
dispersion, as addressed by her, can be attributed to 
evolutionary effects in these regions through the presence of several ionising clusters
with different ages and functional parameters. Furthermore, she found that the
Ne$^{2+}$/O$^{2+}$ ionic ratio decreases with increasing O$^{2+}$/O ratio.\\
 
These observations are in complete disagreement with observations in our
Galaxy. Stellar nucleosynthesis theory 
predicts both Ne and O to be produced solely by massive stars (M {$\geq$} 8M$_{\odot}$),
hence the ionic ratio should be constant for any oxygen abundance.
Observations of  H{\sevensize II} regions in both the Milky Way and the
Magellanic Clouds (Peimbert \& Costero 1969; Pagel {\it et al.} 1978; Simpson 
{\it et al.} 1995) support this constancy. Furthermore, Henry (1990), from
the study of a wide sample of planetary nebula in the Galaxy, the Magellanic
Clouds and M31, finds a perfect 
correlation between Ne and O that holds over a range of more
than an order of magnitude for the abundances of these two elements. These
observations seem to be robust in the sense that planetary nebula are formed from
the ejection of the dying star outer envelope that contains He, C and N. Then, the 
observed constancy in the Ne/O ratio implies that intermediate-mass stars 
(B stars) neither produce nor destroy significant amounts of neon and oxygen 
during their evolution through the HR diagram. Hence this 
analysis could explain why galactic H{\sevensize II} regions show a constant 
Ne/O ratio as only a few OB stars, with the same properties, 
are required to ionise these regions.\\

We have plotted the H$\beta$ equivalent
width {\it versus} the Ne$^{2+}$/O$^{2+}$ ionic ratio (Figure 12) for a sample of 
well studied
GEHR's with Wolf-Rayet stars embedded in them. Evolutionary effects are 
readily apparent 
in NGC 604 (M33) (see D{\'\i}az {\it et al.} 1987). We can see 
a 0.2-0.3 dex difference in the observed EW(H$\beta$) values between the individual 
ionising clusters (604-B and 604-D) and the full aperture integrated region (604). 
As for log(Ne$^{2+}$/O$^{2+}$), there is a 0.1 dex and 0.02 dex difference
between the integrated spectrum and 604B, 604D respectively. Finally,
the derived oxygen abundance for the whole NGC 604 region differs 
in 0.1 dex from both individual regions.
Hence, a non-negligible difference is found
between the individual regions and the fully integrated spectra. Figure 12 shows 
an increase in the Ne$^{2+}$/O$^{2+}$ ionic ratio with decreasing H$\beta$ 
equivalent width, as Masegosa found for her sample. 
But, the main problem is that we cannot ensure that either the
observed regions in the sample are individual regions (i.e. ionised by a single star cluster)
or can be considered representative of the whole emitting volume. Moreover, it could be
possible to sample partial contributions from different regions (i.e. star clusters). All these
possible combinations would yield different values for the EW(H$\beta$), and the slope in
the diagram would change. It is apparent that the EW(H$\beta$) can produce misleading
trends when used as a diagnostic indicator.
As for the ionic ratio, the inferred difference of 0.1 dex in the log(Ne$^{2+}$/O$^{2+}$) 
ratio between 604B and NGC 604 cannot be considered  significative
within the observational errors.

The problem would be solved if we could observe each ionising 
cluster separately. In this case, each 
one could be treated as a single galactic H{\sevensize II} region, and the 
derived properties would be single-valued. This cannot be achieved 
until high spatial resolution spectroscopy is performed in a significant
number of GEHR's.
%

\section{Summary and conclusions}

We have made long-slit spectrophotometric observations in the
optical and near infrared of 14 H{\sevensize II} regions in different spiral
galaxies (NGC 628, NGC 925, NGC 1232 and NGC 1637). For all the sample, a relatively
high oxygen abundance (solar or oversolar) has been reported in the literature (VZ98)
as deduced from empirical calibrations based on the optical oxygen forbidden lines.
These spectrophotometric observations were performed with a wide spectral
coverage and at a resolution high enough to detect and measure both weak 
auroral forbidden lines and Wolf-Rayet features.
Electron temperatures have been derived in order to investigate the ionisation structure
and to derive the chemical composition of the gas in these regions.

For seven of the observed regions, H13, H3, H4, H5 
(NGC 628); CDT1, CDT3 and CDT4 (NGC 1232), we have been able to measure 
reliable 
electron temperatures from different nebular to auroral line intensity
ratios, which allows the derivation of accurate abundances following
standard methods. In particular, the metallicities found
for these regions (except CDT1 in NGC 1232), previously reported to be close to solar, are found to be
lower by factors between 0.2 and 0.5 dex, the latter in the case of H13 and H3.
For the rest of the regions both an empirical calibration 
based on the sulphur emission lines and detailed modelling has been 
used to determine a mean oxygen content.

In the case of CDT1 (NGC 1232), it is the first time that it has been possible 
to derive, in a consistent manner, the mean oxygen content in a high metallicity H{\sevensize II} region
following standard methods. The derived value is 12 + log(O/H) = 8.95, 
though a conservative value of 8.85 can be adopted under ionisation 
structure considerations. This region,
together with one observed in NGC 1637 (CDT1) (its oxygen abundance
derived from detailed modelling), are the only two regions
in the sample that can be considered as of high metallicity.

Our data and those from different empirical calibrations are compared in order to 
investigate the radial abundance gradients in NGC~628 and NGC~1232. It can 
be inferred from this analysis that, except for
 Pilyugin's calibration (Pilyugin 2000), 
which is in fair agreement with our observed values, all of the others 
overestimate, 
by a factor of about 3, the oxygen abundances in relatively low 
excitation H{\sevensize II} regions.

Our observational results show a mean S/O abundance ratio  of {$\pm$} 0.2 dex
around the solar value. Nevertheless, it is important to stress that both regions 
H13 and H3 (NGC~628) show a subsolar 
S/O abundance ratio. Regarding the high metallicity region CDT1 in NGC~1232,
a solar S/O abundance ratio is found, and this value must be considered an 
upper limit for this abundance ratio. This result seems to support the trends, 
previously found 
by D{\'\i}az {\it et al.} (1991) from detailed modelling for the observed 
high metallicity H{\sevensize II} regions in M51.
Regarding the N/O abundance ratio, all the observed regions show undersolar
or solar values within the errors, except CDT1 (NGC 1232) where a direct value 
of -0.81 is obtained implying a secondary contribution of nitrogen 
as metallicity increases.

In the four observed GEHR's (H13, CDT1, CDT3 and CDT4), 
ion-weighted temperatures have been derived from different auroral to nebular line
ratios. Our results show that mean ion-weighted
temperatures from single-ionised atoms (N$^{+}$ and S$^{+}$) are higher than those 
corresponding to twice-ionised species (S$^{2+}$) for electron temperatures 
below 10,000 K. The derived values are in excellent 
agreement with predictions from single-star photoionisation
models (Garnett 1992).
In the case of region H13, the mean value of the four measured electron
temperatures is 10,000 $\pm$ 700 K, and, again, models predict this isothermal 
behaviour at 10,000 K. Moreover, by adopting the new effective collision strengths from
Tayal (1997), several important discrepancies between T(S$^{2+}$) and T(O$^{2+}$)
in other well studied GEHR's can be successfully removed. Therefore, 
Garnett's three-zone model nebula (Garnett 1992) seems to explain 
consistently the temperature stratification in H{\sevensize II} regions.

Wolf-Rayet features have been detected in the four observed GEHR's, and are specially
prominent in region H13. In this case, when
comparing both the emission line spectra and the inferred abundances with those
derived in the nearby region H3, it is confirmed that they closely resemble each other,
despite the presence of Wolf-Rayet stars in region H13. From this result, it can
be concluded that the presence of WR stars in this region does not alter its
ionisation structure. This fact can be understood in terms of the low derived
value for the WR/O star ratio (Castellanos {\it et al.} 2001, Paper II, in preparation).
Another fact that supports the previous conclusion is the isothermal behaviour observed
in H13.

From the calculation of single-star photoionisation models, one striking result
is the derived constancy for the mean effective temperature of the
ionising clusters in the four observed GEHR's, with a value around 35,000 K (Mihalas models).
This result is remarkable given the scatter in the derived 
metal abundances and Wolf-Rayet properties for the observed GEHR's.

\section*{Acknowledgements}

We thank Bernard Pagel for his suggestions which significantly
improved the content of the paper as well as an
anonymous referee for many useful comments.
The WHT is operated in the island of La Palma by the Issac Newton Group 
in the Spanish Observatorio del 
Roque de los Muchachos of the Instituto de Astrof{\'\i}sica de Canarias. We 
would like to thank CAT for awarding observing time.

E.T. is grateful to an IBERDROLA Visiting Professorship 
to UAM during which part of this work was completed.
This work has been partially supported by DGICYT project PB-96-052.


%
%
\begin{figure}
\setcounter{figure}{0}
\centering
\begin{minipage}[c]{10mm}
\centering\psfig{figure=halfa628.epsi,width=8.4cm,height=7.4cm,clip=}
\end{minipage}%
\begin{minipage}[c]{10mm}
\centering\psfig{figure=halfa1232a.epsi,width=8.4cm,height=7.4cm,clip=}
\end{minipage}
\begin{minipage}[c]{10mm}
\centering\psfig{figure=halfa1232b.epsi,width=8.4cm,height=7.4cm,clip=}
\end{minipage}%
\begin{minipage}[c]{10mm}
\centering\psfig{figure=halfa1232c.epsi,width=8.4cm,height=7.4cm,clip=}
\end{minipage}
\begin{minipage}[c]{10mm}
\centering\psfig{figure=halfa925.epsi,width=8.4cm,height=7.4cm,clip=}
\end{minipage}%
\begin{minipage}[c]{10mm}
\centering\psfig{figure=halfa1637.epsi,width=8.4cm,height=7.4cm,clip=}
\end{minipage}
\caption{H$\alpha$ profiles for the observed slit positions. Each figure
include the name of the galaxy, the P.A. of the slit, the name of the observed
regions and their X,Y positions (see also Table 2).}
\end{figure}
%
%
\begin{figure*}
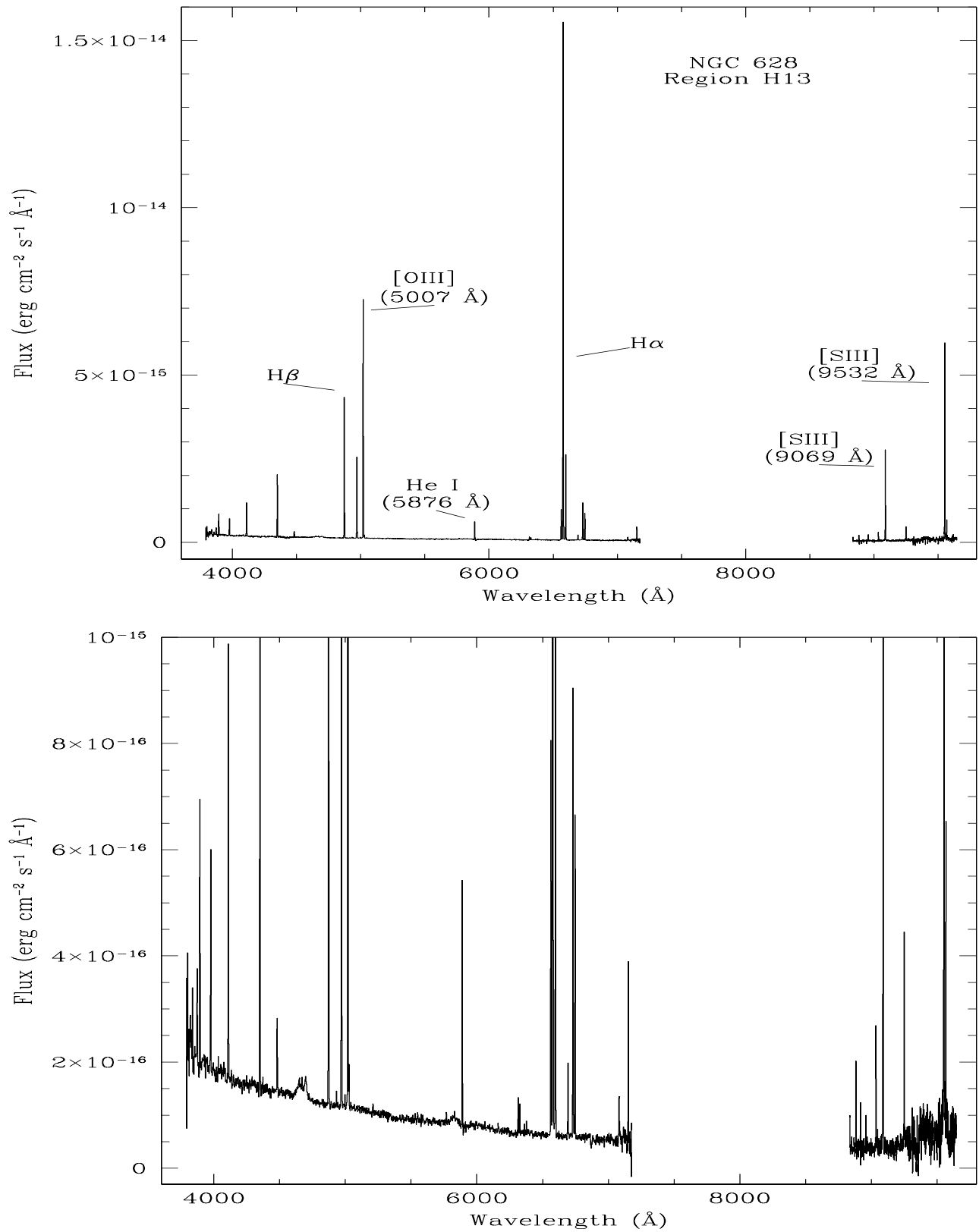

\setcounter{figure}{1}
\begin{minipage}{180mm}
\psfig{figure=h13.epsi,height=10.5cm,width=17cm,clip=}
\vspace{12pt}
\psfig{figure=scaledh13.epsi,height=10.5cm,width=17cm,clip=}
\vspace{12pt}
\caption{Merged spectrum for region H13 in NGC 628, with two intensity scales.}
\end{minipage}
\end{figure*} 
\begin{figure*}
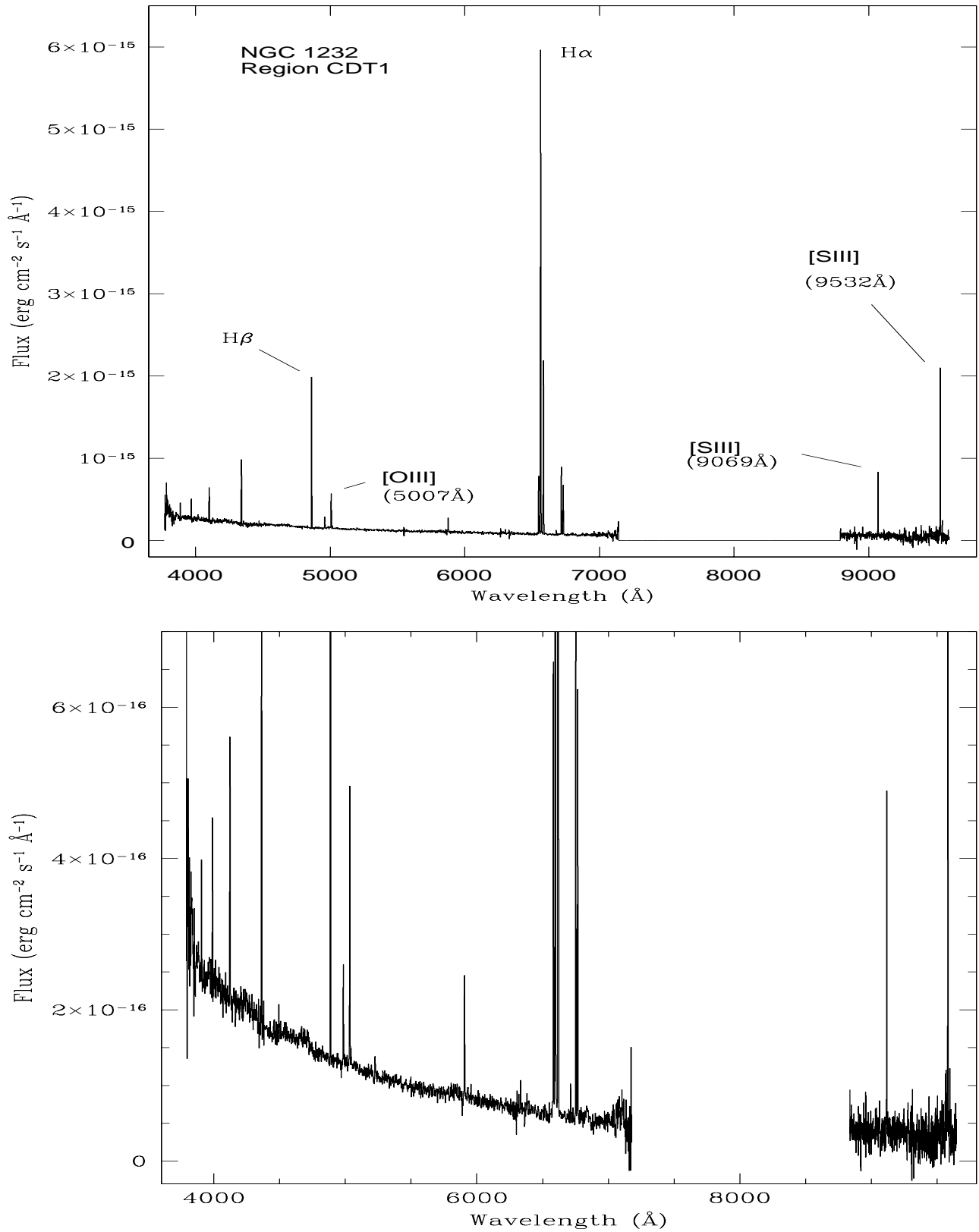
 
\setcounter{figure}{2}
\begin{minipage}{180mm}
\psfig{figure=CDT1.epsi,height=10.5cm,width=17cm,clip=}
\vspace{12pt}
\psfig{figure=scaledCDT1.epsi,height=10.5cm,width=17cm,clip=} 
\vspace{12pt}
\caption{Merged spectrum for the high metallicity region CDT1 in NGC 1232, with two
intensity scales}
\end{minipage}
\end{figure*}
\begin{figure*}
\setcounter{figure}{3}
 \psfig{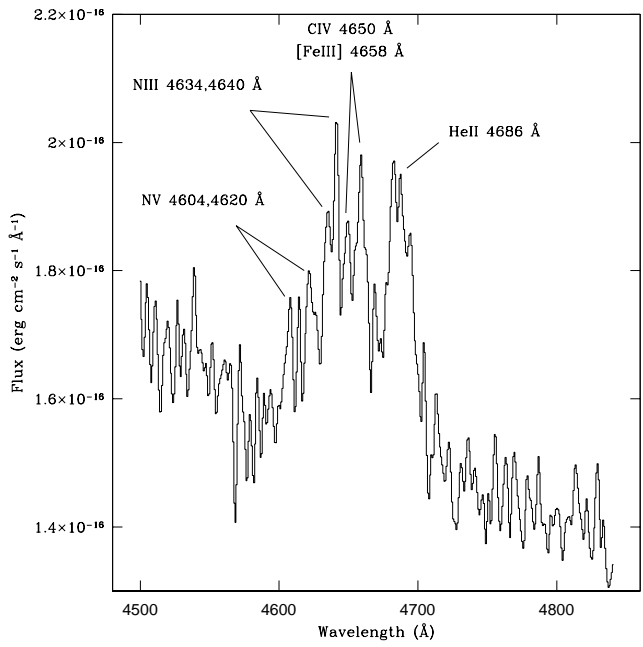}
 \vspace{10pt}
 \psfig{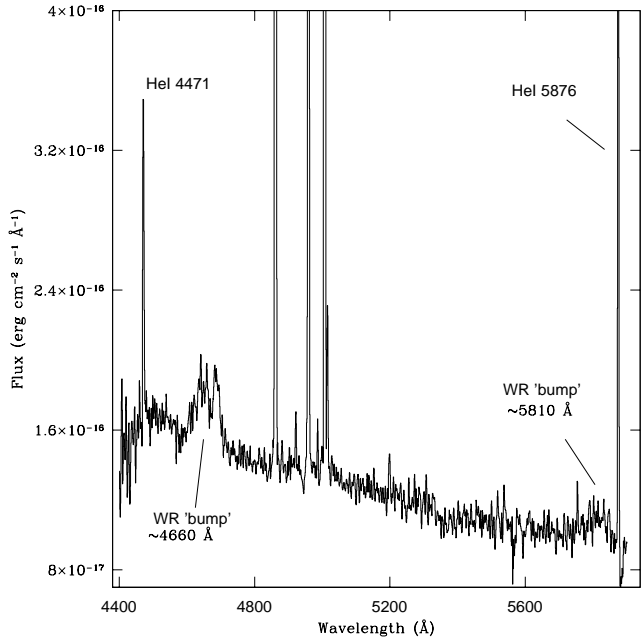}
\caption{Wolf-Rayet features in region H13 (NGC 628).{\it Upper}: blue WR
 'bump'. {\it Bottom}: blue and red WR 'bumps'}
\end{figure*}
\begin{figure*}
\setcounter{figure}{4}
\psfig{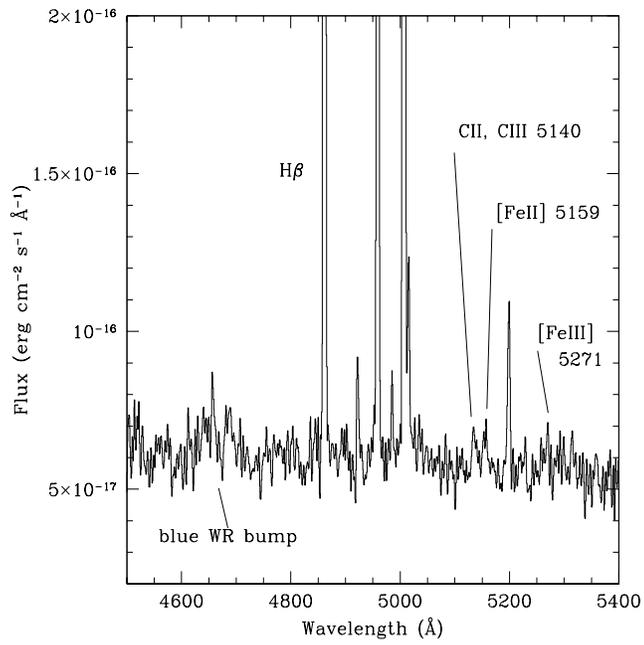}
 \vspace{10pt}
\psfig{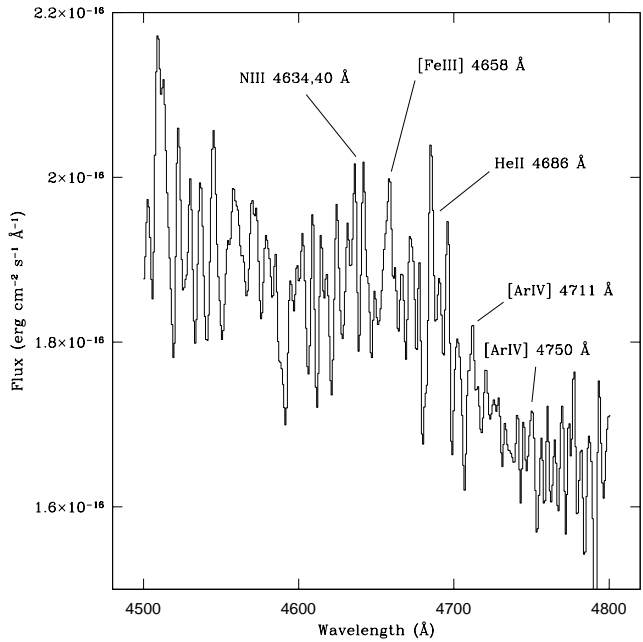}
\caption{NGC 1232.{\it Top}: Region CDT3 with the  blue WR 'bump', C{\sevensize II} and C{\sevensize III} 
(WC stars) and excitation lines. {\it Bottom}: WR bump in the high metallicity region CDT1}
\end{figure*}
%
\begin{figure*}
\setcounter{figure}{5}
 \psfig{figure=628data.epsi,height=8.4cm,width=8.4cm,clip=}
\caption{The oxygen abundance gradient of NGC~628 as derived using
  different empirical calibrations as compared to the data in this
  work. The meaning of the different symbols as explained in text}
\end{figure*}
\clearpage
\begin{figure*}
\setcounter{figure}{6}
 \psfig{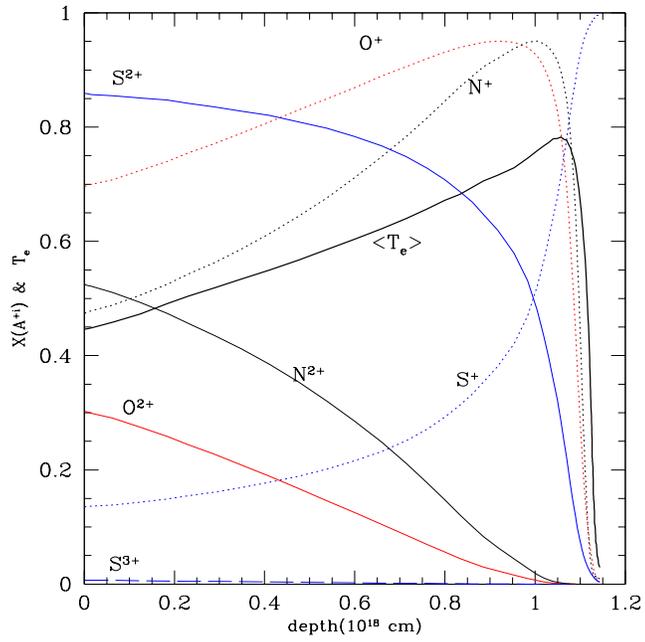}
\caption{The ionisation structure of a model H{\sevensize II} region with the functional
  parameters derived for region CTD1 in NGC~1232}
\end{figure*}
\clearpage
\begin{figure*}
\setcounter{figure}{7}
 \psfig{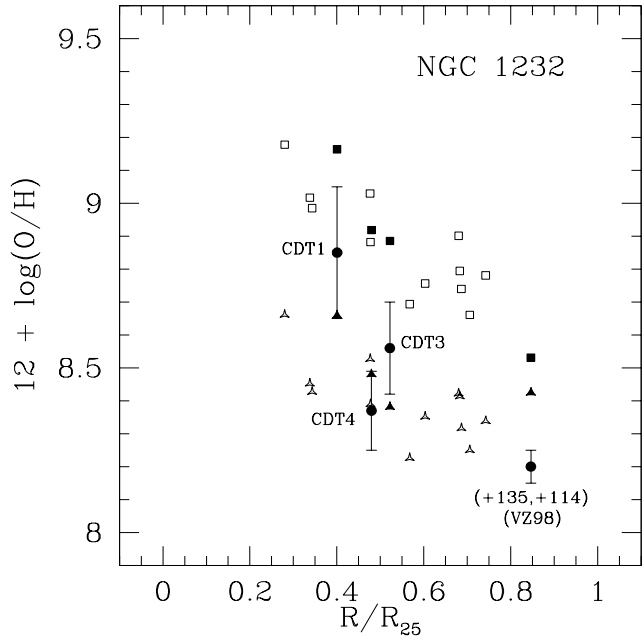}
\caption{The oxygen abundance gradient of NGC~1232 as derived using
  different empirical calibrations as compared to the data in this
  work. The meaning of the different symbols as explained in text.}
\end{figure*}
\clearpage
\begin{figure*}
\setcounter{figure}{8}
 \psfig{figure=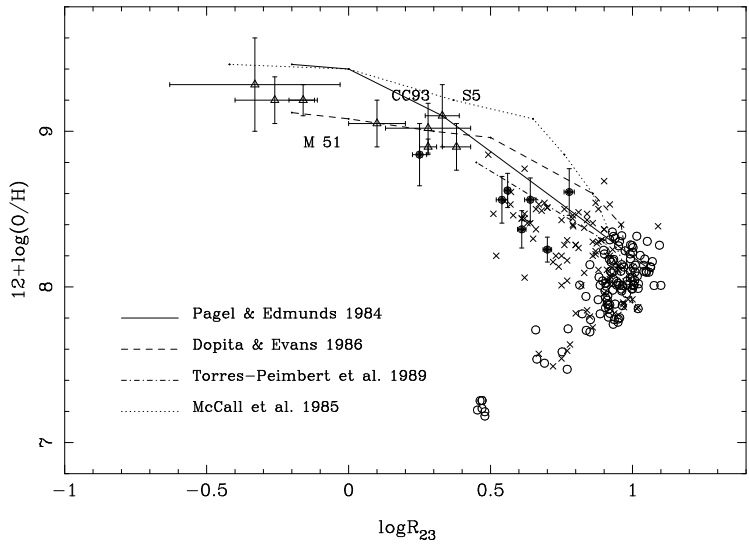,rheight=8.4cm,width=8.4cm,angle=270}
\caption{The oxygen abundance {\em vs} R$_{23}$ diagram. Crosses
  correspond to H{\sevensize II} regions in spiral and irregular galaxies. Open
  circles correspond to H{\sevensize II} galaxies. Open triangles correspond to high
  metallicity H{\sevensize II} regions for which a detailed modelling has been
  performed. Finally, the data corresponding to the present work are
  shown as filled circles. Four different calibrations are shown, as labelled.}
\end{figure*}
\clearpage
\begin{figure*}
\setcounter{figure}{9}
 \psfig{figure=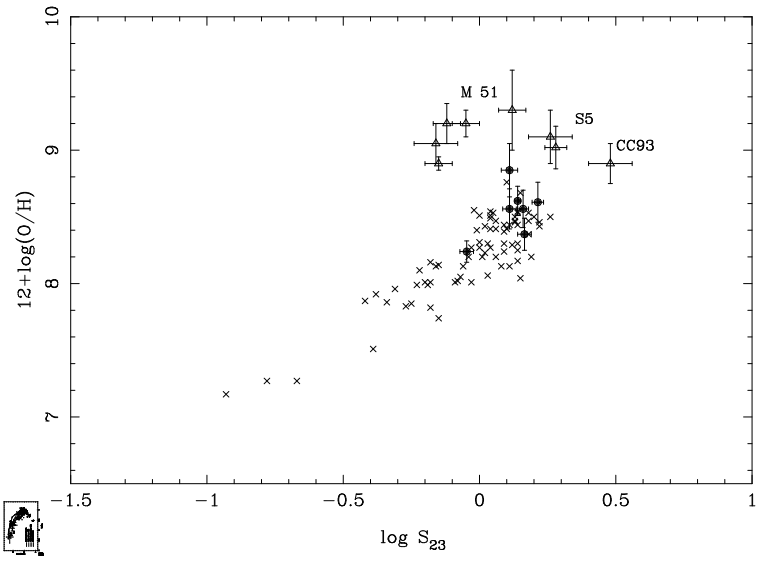,rheight=8.4cm,width=8.4cm,angle=270}
\caption{The oxygen abundance {\em vs} S$_{23}$ diagram. Crosses
  correspond to H{\sevensize II} regions in spiral and irregular galaxies. 
 Open triangles correspond to high
  metallicity H{\sevensize II} regions for which a detailed modelling has been
  performed. Finally, the data corresponding to the present work are
  shown as filled circles.}
\end{figure*}
\clearpage
%
%
%
%
\begin{figure*}
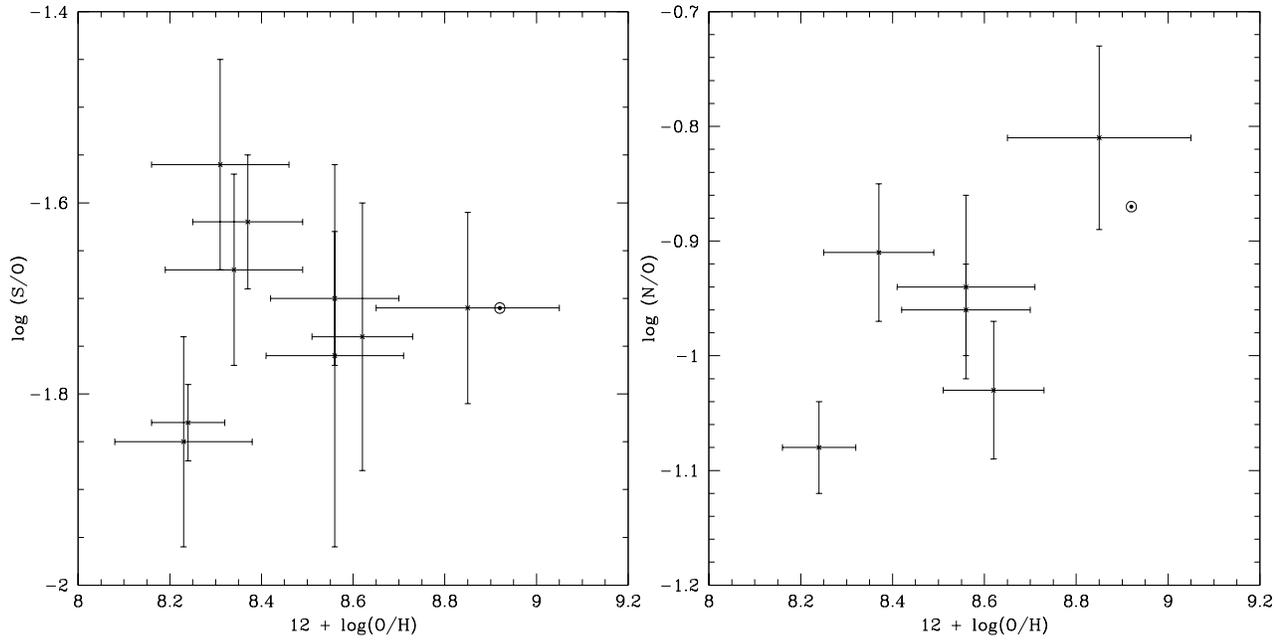

\setcounter{figure}{10}
\centering
\begin{minipage}[c]{15cm}
\centering\psfig{figure=sulfursample2.epsi,height=8.4cm,width=8.4cm,clip=}
\end{minipage}%
\begin{minipage}[c]{15cm}
\centering\psfig{figure=nitrosample.epsi,height=8.4cm,width=8.4cm,clip=}
\end{minipage}
\caption{S/O (left) and N/O (right) abundance ratios as a function of
oxygen abundance in those regions where ion-weighted temperatures have
been  measured. Regions 74C and 69C in NGC 4258 (D{\'\i}az {\it et al.} (2000) are also included. 
$\odot$ stands for the adopted solar abundance ratios.}
\end{figure*}
\clearpage
%
%
%
%
\begin{figure*}
\begin{minipage}[c]{15cm}
\centering
\setcounter{figure}{11}
 \hspace{7cm}
 \psfig{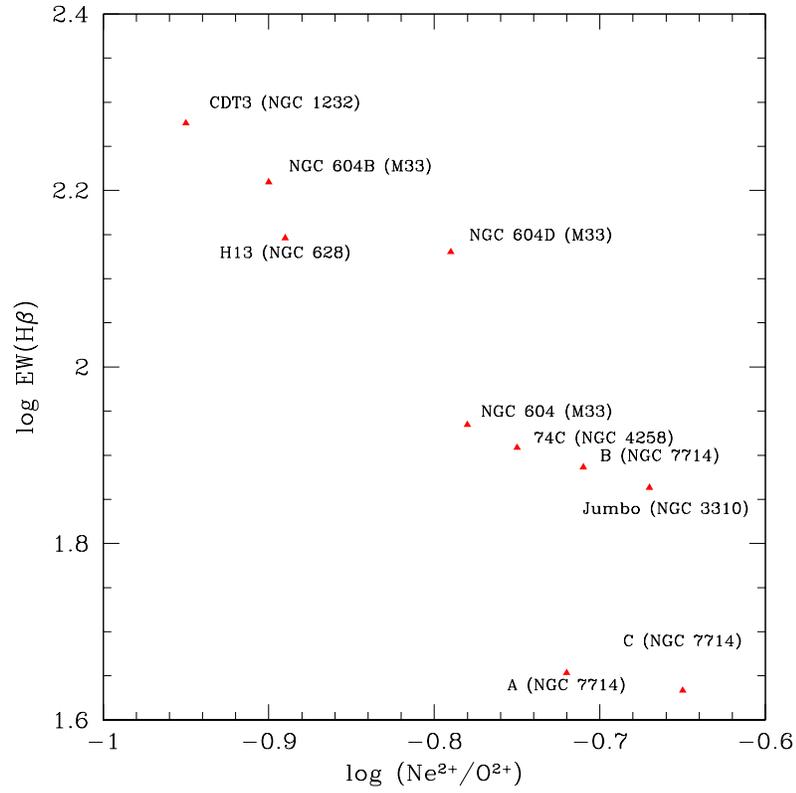}
 \caption{The H$\beta$ equivalent width {\it versus} Ne$^{2+}$/O$^{2+}$ ionic ratio for
a well studied sample of GEHR's.}
\end{minipage}
\end{figure*}
%
\clearpage
\begin{table*}
\setcounter{table}{0}
\centering
 \begin{minipage}{250mm}
 \caption{The galaxy sample}
 \begin{tabular}{@{}lccccc@{}}
\hline
\hline
Property & NGC 628 & NGC 925 & NGC 1232 & NGC 1637 \\
\hline
Type              & .SAS5  & .SXS7 & .SXT5 & .SXT5 \\
Distance (Mpc)    &  7.3    &    8.6    & 21.5      & 7.8 \\
M$_{B}$ \footnote{NGC 628, Sharina {\it et al.} (1996); NGC 925 and NGC 1637, Sohn \& Davidge (1998); NGC 1232, (RC3)}&  -19.5  &    -21.07 & -21.2     & -19.72 \\
i ($\circ$)        &  25     &    58     &  30       &  36 \\
R$_{25}$ ($^{\prime\prime}$) &  314    &    314    & 222       & 120 \\
\hline
 \end{tabular}
 \end{minipage}
\end{table*}
\clearpage
%
%
\begin{table*}
\setcounter{table}{1}
 \begin{minipage}{250mm}
 \caption{Journal of observations}
 \begin{tabular}{@{}lcccccc@{}}
\hline
\hline
Galaxy & P.A. (\degr)\footnote{All the observations made on 1999 November 8/9} & Slit position\footnote{E-W, N-S} & Grating & $\lambda$ range ({\AA}) & Exposure (s) & Mean airmass (sec z) \\
\hline
NGC 628 & 38 & (-086,+186) & R300B & 3800-7200 & 2$\times$1800+1$\times$1200 & 1.050   \\
        & 38 &             & R600R & 8840-9650 & 2$\times$1800+1$\times$1200 &         \\
NGC 925 & 103 & (-008,+000) & R300B & 3800-7200 & 2$\times$1200 & 1.896  \\
        & 103 &             & R600R & 8840-9650 & 2$\times$1200 &        \\
NGC 1232 & 357 & (+059,+078) & R300B & 3800-7200 & 2$\times$1800 & 1.547 \\
         & 357 &             & R600R & 8840-9650 & 2$\times$1800 &       \\
         &  78 & (+004,-101) & R300B & 3800-7200 & 1$\times$1800+1$\times$1200 & 1.596 \\
         &  78 &             & R600R & 8840-9650 & 1$\times$1800+1$\times$1200 &       \\
         &  52 & (+021,+091) & R300B & 3800-7200 & 2$\times$1200 & 1.894 \\
         &  52 &             & R600R & 8840-9650 & 2$\times$1200 &       \\
NGC 1637 &  24 & (+047,-031) & R300B & 3800-7200 & 1800 & 1.504 \\
         &  24 &             & R600R & 8840-9650 & 1800 &       \\
GD 50 (standard) &  &  & R300B & 3800-7200 & 600 & 1.298 \\
                 &  &  & R600R & 8840-9650 & 1200 &     \\
HD 93521 (standard) &  &   & R300B & 3800-7200 & 2 & 1.190 \\
                    &  &  & R600R & 8840-9650 & 60 &       \\
Feige 34 (standard) &  &  & R300B & 3800-7200 & 60 & 1.150 \\
                    &  &  & R600R & 8840-9650 & 600 &      \\
HD 93521 (standard) &  &  & R300B & 3800-7200 & 2 & 1.115 \\
                    &  &  & R600R & 8840-9650 & 60 &       \\
\hline
 \end{tabular}
 \end{minipage}
\end{table*}
\newpage
%
%
\begin{table*} 
\setcounter{table}{2}  
\begin{minipage}{148mm}   
\caption{NGC 628. Reddening corrected line intensities for P.A.=
38$^{\circ}$}   \scriptsize{   \begin{tabular}{@{}llcccc@{}}   

\hline   
\hline
   Region & & H13 (-085,+193) & H3 (-068,+214) & H4 (-066,+217) & H5 (-064,
+218)\\      ~    & &     &     \\     Line   & &     &     \\   \hline  3727
(VZ98) & [OII]        & 2960 $\pm$ 110 & -- & -- & --\\  3835 &  H9+HeII      
     &   52 $\pm$  2 &  --& -- & --           \\  3869 & [NeIII]             &
  60 $\pm$  8 &   92 $\pm$ 11 & -- & -- \\  3869 (VZ98) & [NeIII]      &   70
$\pm$  3 & -- & -- & -- \\  3888 &  H8+HeI             &  157 $\pm$  5 &  161
$\pm$ 10 & 131 $\pm$ 7 & 143 $\pm$ 9 \\  3970 & [NeIII]+H$\epsilon$ &  156
$\pm$  2 &  177 $\pm$  6 & 113 $\pm$ 4 & 126 $\pm$ 4 \\  4068 & [SII]         
     &   17 $\pm$  2 &  -- & -- & --          \\  4076 & [SII]               &
  12 $\pm$  1 &  -- & -- & --          \\   4102 &  H$\delta$          &  246
$\pm$ 10 &  221 $\pm$  5 & 215 $\pm$ 7 & 208 $\pm$ 8 \\  4340 &  H$\gamma$    
     &  458 $\pm$  8 &  462 $\pm$  7 & 427 $\pm$ 15 & 430 $\pm$ 10 \\  4363 &
[OIII]              &   10 $\pm$  2 &  16 $\pm$ 2 & -- & -- \\  4471 &  HeI   
            &   44 $\pm$  1 &  36 $\pm$  1 & -- & -- \\  4658 &  [FeIII]      
     &    8 $\pm$  1 &  -- & -- & -- \\  4711 &  [ArIV]             &    3
$\pm$  1 &  -- & -- & -- \\  4861 &  H$\beta$           & 1000 $\pm$ 20 & 1000
$\pm$ 10 & 1000 $\pm$ 30 & 1000 $\pm$ 30 \\   4881 & [FeIII]             &   
3 $\pm$  1 &  -- & -- & -- \\  4922 &  HeI                &    7 $\pm$  1 & 
-- & -- & -- \\  4959 & [OIII]              &  526 $\pm$  5 &  510 $\pm$  6 &
166 $\pm$ 4 & 175 $\pm$ 4 \\  4987 & [FeIII]             &    7 $\pm$  1 &  --
& -- & -- \\  5007 & [OIII]              & 1547 $\pm$ 10 & 1596
$\pm$ 12 & 488 $\pm$ 8 & 519 $\pm$ 8 \\  4959+5007 & [OIII]         & 2073
$\pm$ 15 &  -- & -- & -- \\  4959+5007 (VZ98) & [OIII]  & 2068 $\pm$ 45 &  --
& -- & -- \\  5012 &  HeI                &   26 $\pm$  1 &  -- & 21 $\pm$  2 &
22 $\pm$ 2  \\  5200 & [NI]                &    6 $\pm$  1 &   14 $\pm$  1 &
14 $\pm$ 1 & 11 $\pm$ 1\\  5517 & [ClIII]             &    4 $\pm$  1 &  -- &
-- & -- \\  5537 & [ClIII]             &    5 $\pm$  1 &  -- & -- & -- \\ 
5755 & [NII]               &    5 $\pm$  1 &  -- & -- & -- \\  5876 &  HeI    
           &   98 $\pm$  8 &  103 $\pm$  3 & 46 $\pm$ 4 & 45 $\pm$ 3 \\   6300
& [OI]                &   18 $\pm$  1 &   38 $\pm$  4 & 26 $\pm$ 3 & 30 $\pm$
5 \\   6300 (VZ98)
& [OI]                &   18 $\pm$  1 & -- & -- & -- \\  6312 & [SIII]              &   12 $\pm$  1 &   11 $\pm$  1 &  5 $\pm$ 1
&  7 $\pm$ 2 \\6312 (VZ98) & [SIII]              &   10 $\pm$  1 &  -- &  -- & -- \\  6364 & [OI]                &    6 $\pm$  1 &    9 $\pm$  2 &
-- & -- \\  6548 & [NII]               &  160 $\pm$  3 &  162 $\pm$  6 & 205
$\pm$ 15 & 205 $\pm$ 17 \\  6563 &  H$\alpha$          & 2860 $\pm$ 30 & 2860
$\pm$ 100 & 2880 $\pm$ 180 & 2870 $\pm$ 170 \\  6584 & [NII]               & 
496 $\pm$  8 &  470 $\pm$ 20 & 603 $\pm$ 40 & 603 $\pm$ 43 \\  6548+6584 &
[NII]          &  656 $\pm$ 11 &  -- & -- & -- \\  6548+6584 (VZ98) & [NII]  
&  686 $\pm$ 24 &  -- & -- & -- \\  6678 &  HeI                &   27 $\pm$  2
&   26 $\pm$  2 & 19 $\pm$ 2 & 18 $\pm$ 2 \\  6717 & [SII]               & 
204 $\pm$  5 &  270 $\pm$ 13 & 309 $\pm$ 20 & 333 $\pm$ 24 \\  6731 & [SII]   
           &  151 $\pm$  4 &  186 $\pm$  9 & 215 $\pm$ 15 & 227 $\pm$ 17 \\ 
6717+6731 & [SII]          &  355 $\pm$  9 &  -- & -- & -- \\  6717+6731
(VZ98) & [SII]   &  360 $\pm$ 12 &  -- & -- & -- \\  7065 &  HeI              
 &   17 $\pm$  2 &  --  & -- & --         \\  8863 &  P11                &   
8 $\pm$  2 &    9 $\pm$ 2 & 10 $\pm$ 2 & 11 $\pm$ 1 \\  9014 &  P10           
    &   11 $\pm$  3 &   12 $\pm$  3 & 13 $\pm$ 2 & 15 $\pm$ 2 \\  9069 &
[SIII]              &  168 $\pm$ 10 &  136 $\pm$ 10 & 147 $\pm$ 12 & 149 $\pm$
10 \\   9229 &  P9                 &   26 $\pm$  1 &   24 $\pm$  2 & 26 $\pm$
1 & 26 $\pm$ 2 \\  9532 & [SIII]              &  374 $\pm$ 15 &  351 $\pm$ 25
& 385 $\pm$ 23 & 399 $\pm$ 20\\   9546 &  P8                 &   34 $\pm$  2 &
  37 $\pm$ 3 & 38 $\pm$ 2 & 34 $\pm$ 3 \\   c(H$\beta$) &              &  
0.29 $\pm$ 0.01  & 0.44 $\pm$ 0.01 & 0.48 $\pm$ 0.03 & 0.47 $\pm$ 0.03 \\ 
{F(H$\alpha$)  \footnote{10$^{-16}$ erg cm$^{-2}$ s$^{-1}$, corrected for
reddening}}& & 1039 & 187 & 109 & 126 \\  EW(H$\beta$)({\AA})& & 140 & 231 &
207 & 215 \\  EW(H$\beta$)({\AA}) (VZ98)& & 152 & -- & -- & -- \\  \hline
\end{tabular} }
\end{minipage}
\end{table*}
\newpage
\begin{table*}
\setcounter{table}{3}
 \begin{minipage}{148mm}
  \caption{NGC 1232. Reddening corrected line fluxes for P.A.= 357$^{\circ}$, 78$^{\circ}$ and 52$^{\circ}$}
  \scriptsize{
  \begin{tabular}{@{}llcccccc@{}}
  \hline
  \hline
   Region & & CDT1 (+059,+058) & CDT2 (+057,+098) & CDT3 (+004,-101) & CDT4 (+021,+091) & CDT5 (-019,+060)  \\
   Line   & &       &       &     &     &   \\ 
  \hline
 3727 (VZ98) & [OII] & 1490 $\pm$ 70 & 4180 $\pm$ 220 & 3180 $\pm$ 150 & 2530 $\pm$ 120 & -- \\ 
 3869 & [NeIII] & -- & -- & 25 $\pm$ 3 & -- & -- \\
 3888 &  H8+HeI & 128 $\pm$ 5 & -- & 92 $\pm$ 4 & 187 $\pm$ 10 & -- \\
 3970 & [NeIII]+H$\epsilon$ & 128 $\pm$ 4 & -- & 96 $\pm$ 5 & 217 $\pm$ 10 & 231 $\pm$ 3 \\ 
 4068,76 & [SII]   &             &              & 48 $\pm$ 6  &    &   \\ 
 4102 &  H$\delta$ & 242 $\pm$ 6 & 256 $\pm$ 16 & 184 $\pm$ 7 & 268 $\pm$ 5 & 261 $\pm$ 5   \\
 4340 &  H$\gamma$ & 470 $\pm$ 7 & 510 $\pm$ 20 & 374 $\pm$ 10 & 466 $\pm$ 7 & 489 $\pm$ 15  \\
 4471 &  HeI & 19 $\pm$ 1 & 65 $\pm$ 1 & 32 $\pm$ 2 & 37 $\pm$ 3 & -- \\
 4861 &  H$\beta$ & 1000 $\pm$ 20 & 1000 $\pm$ 20 & 1000 $\pm$ 12 & 1000 $\pm$ 10 & 1000 $\pm$ 20  \\
 4922 &  HeI & -- & -- & 7 $\pm$ 1 & 10 $\pm$ 2 & -- \\ 
 4959 & [OIII] & 77 $\pm$ 2 & 456 $\pm$ 8 & 284 $\pm$ 6 & 393 $\pm$ 3 & 63 $\pm$ 2  \\
 4987 & [FeIII] & -- & -- & 6 $\pm$ 2 & -- & -- \\
 5007 & [OIII] & 229 $\pm$ 4 & 1350 $\pm$ 20 & 842 $\pm$ 13 & 1162 $\pm$ 8 & 200 $\pm$ 4 \\
 4959+5007 & [OIII] & 306 $\pm$ 6 & 1806 $\pm$ 28 & 1126 $\pm$ 19 & 1555 $\pm$ 11 & 263 $\pm$ 6 \\
 4959+5007 (VZ98) & [OIII] & 300 $\pm$ 10 & 1590 $\pm$ 55 & 979 $\pm$ 30 & 1349 $\pm$ 41 & -- \\
 5012 &  HeI & 10 $\pm$ 1 & -- & 24 $\pm$ 2 & 17 $\pm$ 1 & -- \\
 5200 & [NI] & 18 $\pm$ 1 & -- & 22 $\pm$ 2 & 41 $\pm$ 2 & -- \\
 5755 & [NII] & 4 $\pm$ 1 & -- & 8 $\pm$ 2 & 8 $\pm$ 1 & -- \\
 5876 &  HeI & 84 $\pm$ 2 & 75 $\pm$ 8 & 120 $\pm$ 6 & 122 $\pm$ 5 & 77 $\pm$ 2 \\
 6300 & [OI] & 25 $\pm$ 2 & 111 $\pm$ 15 & 43 $\pm$ 3 & 35 $\pm$ 4 & 30 $\pm$ 5 \\
 6300 (VZ98) & [OI] & 19 $\pm$ 1 & 131 $\pm$ 9 & 45 $\pm$ 2 & 32 $\pm$ 2 & -- \\
 6312 & [SIII] & 2 $\pm$ 1 & -- & 8 $\pm$ 2 & 13 $\pm$ 2 & --           \\
 6312 (VZ98) & [SIII] & 2 $\pm$ 1 & 18 $\pm$ 5 & 7 $\pm$ 1 & 7 $\pm$ 1 & --           \\
 6364 & [OI] & 9 $\pm$ 1 & -- & 15 $\pm$ 2 & 11 $\pm$ 2 & -- \\
 6548 & [NII] & 328 $\pm$ 8 & 231 $\pm$ 15 & 290 $\pm$ 10 & 252 $\pm$ 14 & 293 $\pm$ 18  \\
 6563 &  H$\alpha$ & 2870 $\pm$ 50 & 2870 $\pm$ 100 & 2870 $\pm$ 150 & 2870 $\pm$ 20 & 2870 $\pm$ 90 \\
 6584 & [NII] & 1040 $\pm$ 21 & 717 $\pm$ 30 & 892 $\pm$ 30 & 758 $\pm$ 40 & 961 $\pm$ 40  \\
 6548+6584 & [NII] & 1368 $\pm$ 29 & 948 $\pm$ 45 & 1182 $\pm$ 40 & 1010 $\pm$ 54 & 1254 $\pm$ 58 \\
 6548+6584 (VZ98) & [NII] & 1349 $\pm$ 54 & 1030 $\pm$ 42 & 1203 $\pm$ 48 & 1018 $\pm$ 41 & -- \\
 6678 &  HeI & 19 $\pm$ 1 & -- & 31 $\pm$ 3 & 29 $\pm$ 4 & --   \\
 6717 & [SII] & 357 $\pm$ 8 & 537 $\pm$ 15 & 333 $\pm$ 10 & 307 $\pm$ 20 & 400 $\pm$ 14  \\
 6731 & [SII] & 274 $\pm$ 6 & 372 $\pm$ 10 & 272 $\pm$ 9 & 233 $\pm$ 15 & 295 $\pm$ 10  \\
 6717+6731 & [SII] & 631 $\pm$ 14 & 909 $\pm$ 25 & 605 $\pm$ 19 & 540 $\pm$ 35 & 695 $\pm$ 24 \\
 6717+6731 (VZ98) & [SII] & 663 $\pm$ 25 & 1028 $\pm$ 42 & 728 $\pm$ 27 & 580 $\pm$ 21 & --\\
 8863 &  P11 & -- & -- & 14 $\pm$ 2 & 14 $\pm$ 4 & -- \\
 9014 &  P10 & -- & -- & 19 $\pm$ 3 & 19 $\pm$ 4 & -- \\
 9069 & [SIII] & 197 $\pm$ 13 & 191 $\pm$ 15 & 229 $\pm$ 10 & 249 $\pm$ 10 & -- \\
 9229 &  P9 & 24 $\pm$ 2 & -- & 26 $\pm$ 2 & 25 $\pm$ 3 & -- \\
 9532 & [SIII] & 455 $\pm$ 34 & 535 $\pm$ 40 & 614 $\pm$ 30 & 665 $\pm$ 20 & -- \\
 9546 &  P8 & 37 $\pm$ 2 & -- & 30 $\pm$ 6 & 34 $\pm$ 3 & -- \\
 c(H$\beta$)&  & 0.25 $\pm$ 0.01 & 0.32 $\pm$ 0.01 & 0.32 $\pm$ 0.03 & 0.62 $\pm$ 0.02 & 0.31 $\pm$ 0.02   \\
 {F(H$\alpha$)
 \footnote{10$^{-16}$ erg cm$^{-2}$ s$^{-1}$, corrected for reddening}}&  & 376 & 41 & 767 & 871 & 114  \\
 EW(H$\beta$)({\AA}) & & 48 & 84 & 189 & 138 & 63 \\
 EW(H$\beta$)({\AA}) (VZ98) & & 63 & 86 & 191 & 113 & -- \\
\hline
\end{tabular}
}
\end{minipage}
\end{table*}
\newpage
%
%
%
%
\begin{table*}
\setcounter{table}{4}
 \begin{minipage}{148mm}
  \caption{NGC 925. Reddening corrected line fluxes for P.A.= 103$^{\circ}$}
  \footnotesize{
  \begin{tabular}{@{}llccccc@{}}
  \hline
  \hline
   Region & & CDT1 (-008,+000) & CDT2 (-005,+000) & CDT3 (+010,-004) & CDT4 (+042,-011)   \\
   Line   & &       &       &     &     \\ 
  \hline
 3727 (VZ98) & [OII] & 2880 $\pm$ 110 & 2370 $\pm$ 90 & 2940 $\pm$ 120 & 2820 $\pm$ 120 \\
 3888 &  H8+HeI & 177 $\pm$ 12 & -- & -- &  -- \\
 3970 & [NeIII]+H$\epsilon$ & 134 $\pm$ 6 & -- & -- & 150 $\pm$ 17  \\ 
 4102 &  H$\delta$ & 215 $\pm$ 20 & --  & -- & 256 $\pm$ 20   \\
 4340 &  H$\gamma$ & 479 $\pm$ 23 & 430 $\pm$ 20 & 468 $\pm$ 40 & 463 $\pm$ 9  \\
 4861 &  H$\beta$ & 1000 $\pm$ 13 & 1000 $\pm$ 20 & 1000 $\pm$ 60 & 1000 $\pm$ 16  \\
 4959 & [OIII] & 301 $\pm$ 7 & 322 $\pm$ 12 & 315 $\pm$ 13 & 239 $\pm$ 4   \\
 5007 & [OIII] & 849 $\pm$ 14 & 892 $\pm$ 24 & 967 $\pm$ 36 & 737 $\pm$ 10  \\
 4959+5007 & [OIII] & 1150 $\pm$ 21 & 1214 $\pm$ 36 & 1282 $\pm$ 49 & 976 $\pm$ 14 \\
 4959+5007 (VZ98) & [OIII] & 915 $\pm$ 23 & 698 $\pm$ 19 & 826 $\pm$ 24 & 1275 $\pm$ 35 \\ 
 5200 & [NI] & -- & -- & --  & 36 $\pm$ 4  \\
 5876 &  HeI & -- & -- & -- & 64 $\pm$ 2 \\
 6300 & [OI] & 41 $\pm$ 4 & 94 $\pm$ 9 & 90 $\pm$ 10 & 39 $\pm$ 2  \\
 6300 (VZ98) & [OI] & 43 $\pm$ 4 & -- & 59 $\pm$ 5 & -- \\
 6548 & [NII] & 203 $\pm$ 11 & 280 $\pm$ 20 & 242 $\pm$ 18 & 196 $\pm$ 7  \\
 6563 &  H$\alpha$ & 2800 $\pm$ 130 & 2890 $\pm$ 190 & 2880 $\pm$ 190 & 2850 $\pm$ 80  \\
 6584 & [NII] & 568 $\pm$ 28 & 723 $\pm$ 40 & 661 $\pm$ 30 & 593 $\pm$ 18  \\
 6548+6584 & [NII] & 771 $\pm$ 39 & 1003 $\pm$ 60 & 903 $\pm$ 48 & 789 $\pm$ 25 \\
 6548+6584 (VZ98) & [NII] & 866 $\pm$ 32 & 899 $\pm$ 34 & 919 $\pm$ 35 & 758 $\pm$ 30 \\
 6678 &  HeI & 28 $\pm$ 3 & 16: & -- & 25 $\pm$ 2   \\
 6717 & [SII] & 577 $\pm$ 30 & 787 $\pm$ 40 & 672 $\pm$ 30 & 437 $\pm$ 15  \\
 6731 & [SII] & 391 $\pm$ 20 & 498 $\pm$ 25 & 449 $\pm$ 22 & 298 $\pm$ 11  \\
 6717+6731 & [SII] & 968 $\pm$ 50 & 1285 $\pm$ 65 & 1121 $\pm$ 52 & 735 $\pm$ 26 \\
 6717+6731 (VZ98) & [SII] & 873 $\pm$ 30 & 931 $\pm$ 32 & 945 $\pm$ 34 & 751 $\pm$ 29 \\ 
 9069 & [SIII] & 125 $\pm$ 9 & 198 $\pm$ 15 & 83 $\pm$ 15 & 160 $\pm$ 10  \\
 9229 &  P9 & -- & -- & -- & 28 $\pm$ 3  \\
 9532 & [SIII] & 359 $\pm$ 25 & 480 $\pm$ 30 & 222 $\pm$ 30 & 342 $\pm$ 25 \\
 c(H$\beta$)&  & 0: & 0.23 $\pm$ 0.03 & 0.19 $\pm$ 0.02 & 0.32 $\pm$ 0.01   \\
 {F(H$\alpha$)
 \footnote{10$^{-16}$ erg cm$^{-2}$ s$^{-1}$, corrected for reddening}}&  & 103 & 66 & 93 & 143  \\
 EW(H$\beta$)({\AA}) & & 22 & 6 & 15 & 47 \\
 EW(H$\beta$)({\AA}) (VZ98) & & 17 & 13 & 15 & 44 \\
\hline
\end{tabular}
}
\end{minipage}
\end{table*}
\newpage
%
%

%
%
\begin{table*}
\setcounter{table}{5}
 \begin{minipage}{148mm}
  \caption{NGC 1637. Reddening corrected line fluxes for P.A.= 24$^{\circ}$}
  \centering
  \small{
  \begin{tabular}{@{}llc@{}}
  \hline
  \hline
   Region & & CDT1 (+047,-031) \\
   Line   & &  \\ 
  \hline
 3727 (VZ98) & [OII] & 1170 $\pm$ 60 \\
 4102 &  H$\delta$ & 220 $\pm$ 10  \\
 4340 &  H$\gamma$ & 455 $\pm$ 7   \\
 4861 &  H$\beta$ & 1000 $\pm$ 10  \\
 4959 & [OIII] & 29 $\pm$ 2    \\
 5007 & [OIII] & 84 $\pm$ 10   \\
 4959+5007 & [OIII] & 113 $\pm$ 12 \\
 4959+5007 (VZ98) & [OIII] & 137 $\pm$ 7 \\
 5876 &  HeI & 71 $\pm$ 10 \\
 6300 & [OI] & 32 $\pm$ 6   \\
 6548 & [NII] & 315 $\pm$ 22   \\
 6563 &  H$\alpha$ & 2840 $\pm$ 30   \\
 6584 & [NII] & 1037 $\pm$ 60   \\
 6548+6584 & [NII] & 1352 $\pm$ 82 \\
 6548+6584 (VZ98) & [NII] & 1304 $\pm$ 54 \\
 6717 & [SII] & 335 $\pm$ 20   \\
 6731 & [SII] & 249 $\pm$ 15   \\
 6717+6731 & [SII] & 584 $\pm$ 35 \\
 6717+6731 (VZ98) & [SII] & 534 $\pm$ 21 \\  
 9069 & [SIII] & 99 $\pm$ 9   \\
 9532 & [SIII] & 247 $\pm$ 15  \\
 c(H$\beta$)&  & 0.52 $\pm$ 0.02  \\
 {F(H$\alpha$)
 \footnote{10$^{-16}$ erg cm$^{-2}$ s$^{-1}$, corrected for reddening}}&  & 243  \\
 EW(H$\beta$)({\AA}) & & 74 \\
 EW(H$\beta$)({\AA}) (VZ98) & & 76 \\
\hline
\end{tabular}
}
\end{minipage}
\end{table*}
\newpage
%
%
\begin{table*}
\setcounter{table}{6}
 \begin{minipage}{200mm}
 \caption{ Derived physical conditions of the gas in the observed H{\sevensize II} regions }
 \footnotesize{
 \begin{tabular}{lccccc}
\hline
 & & NGC~628 & & \\
\hline
\hline
 Parameter  & H13 & H3 & H4 & H5 \\
            &     &     &       &  \\
 n$_{e}$    & 80  & $\leq$ 40 & $\leq$ 40 & $\leq$ 40 \\
 $<$log U$>$ & -2.78 $\pm$ 0.10 & -2.92 $\pm$ 0.15 & -2.95 $\pm$ 0.15 & -2.97 $\pm$ 0.15 \\
 t(S$^{2+}$) & 1.02 $\pm$ 0.03 & 1.03 $\pm$ 0.05 & 0.74 $\pm$ 0.04  & 0.82 $\pm$ 0.06 \\
 t(O$^{2+}$) & 0.98 $\pm$ 0.05 & 1.17 $\pm$ 0.10 & 0.69 & 0.79 \\
 t(S$^{+})$  & 0.99 $\pm$ 0.06 & -- & -- & -- \\
 t(N$^{+})$  & 0.90 $\pm$ 0.06 & -- & -- & -- \\
 t(O$^{+})$  & 0.99 & 1.10 & 0.79 & 0.85 \\
 12 + log(O$^{2+}$/H$^{+}$) & 7.77 $\pm$ 0.07 & 7.53 $\pm$ 0.10 & 7.89 $\pm$ 0.13 & 7.66 $\pm$ 0.15 \\
 12 + log(O$^{+}$/H$^{+}$)  & 8.06 $\pm$ 0.09 & 8.14 $\pm$ 0.16 & 8.10 $\pm$ 0.15 & 8.24 $\pm$ 0.15 \\
 12 + log(O/H)              & 8.24 $\pm$ 0.08 & 8.23 $\pm$ 0.15 & 8.31 $\pm$ 0.15 & 8.34 $\pm$ 0.15 \\
 12 + log(S$^{2+}$/H$^{+}$) & 6.25 $\pm$ 0.02 & 6.20 $\pm$ 0.03 & 6.52 $\pm$ 0.04 & 6.43 $\pm$ 0.05 \\
 12 + log(S$^{+}$/H$^{+}$)  & 5.91 $\pm$ 0.05 & 5.92 $\pm$ 0.06 & 6.35 $\pm$ 0.05 & 6.29 $\pm$ 0.05 \\
 12 + log(S/H)              & 6.41 $\pm$ 0.04 & 6.38 $\pm$ 0.05 & 6.75 $\pm$ 0.04 & 6.67 $\pm$ 0.05 \\
 log(N/O)                   &-1.08 $\pm$ 0.04 & -1.14 $\pm$ 0.10 & -0.94 $\pm$ 0.10 & -1.07 $\pm$ 0.10 \\
 log(S/O)                   &-1.83 $\pm$ 0.04 &-1.85 $\pm$ 0.11 &-1.56 $\pm$ 0.11 &-1.67 $\pm$ 0.10 \\
 log(Ne$^{2+}$/O$^{2+}$)    &-0.89 $\pm$ 0.04 &-0.83 $\pm$ 0.04 & -- & -- \\
 He$^{+}$/H$^{+}$           &0.077 $\pm$ 0.005 &0.074 $\pm$ 0.005& -- & -- \\
 ICF(He)                    & 1.19 & 1.25 & -- & -- \\
 He/H                       & 0.093 $\pm$ 0.006 &0.093 $\pm$ 0.006& -- & -- \\
\hline
 & & NGC~1232 & & \\
\hline
\hline

 Parameter  & CDT1 & CDT2 & CDT3 & CDT4 \\
            &     &     &       &  \\
 n$_{e}$    & 130  & $\leq$ 40 & 223 & 118 \\
 $<$log U$>$ & -2.95 $\pm$ 0.20 & -2.95 $\pm$ 0.15 & -2.72 $\pm$ 0.10 & -2.72 $\pm$ 0.10 \\
 t(S$^{2+}$) & 0.54 $\pm$ 0.05 & -- & 0.74 $\pm$ 0.05  & 0.87 $\pm$ 0.04 \\
 t(O$^{2+}$) & 0.45 & -- & 0.69 & 0.84 \\
 t(S$^{+})$  & -- & -- & 0.90 $\pm$ 0.06 & -- \\
 t(N$^{+})$  & 0.67 $\pm$ 0.05 & -- & 0.86 $\pm$ 0.06 & 0.90 $\pm$ 0.06 \\
 t(O$^{+})$  & 0.62 & -- & 0.79 & 0.89 \\
$<t>_{adop}$   & -- & 0.81 & -- & --     \\ 
 12 + log(O$^{2+}$/H$^{+}$) & 8.63 $\pm$ 0.30 & 8.03 & 8.13 $\pm$ 0.15 & 7.90 $\pm$ 0.10 \\
 12 + log(O$^{+}$/H$^{+}$)  & 8.66 $\pm$ 0.12 & 8.46 & 8.37 $\pm$ 0.12 & 8.20 $\pm$ 0.13 \\
 12 + log(O/H)              & 8.95 $\pm$ 0.20 & 8.61 $\pm$ 0.15 & 8.56 $\pm$ 0.14 & 8.37 $\pm$ 0.12 \\
 12 + log(S$^{2+}$/H$^{+}$) & 6.97 $\pm$ 0.09 & 6.53 & 6.72 $\pm$ 0.06 & 6.60 $\pm$ 0.04 \\
 12 + log(S$^{+}$/H$^{+}$)  & 6.67 $\pm$ 0.11 & 6.47 & 6.30 $\pm$ 0.11 & 6.21 $\pm$ 0.06 \\
 12 + log(S/H)              & 7.14 $\pm$ 0.10 & 6.80 & 6.86 $\pm$ 0.07 & 6.75 $\pm$ 0.05 \\
 log(N/O)                   &-0.81 $\pm$ 0.08 & -1.15 &-0.96 $\pm$ 0.04 &-0.91 $\pm$ 0.06 \\
 log(S/O)                   &-1.81 $\pm$ 0.10 & -1.81 &-1.70 $\pm$ 0.07 &-1.62 $\pm$ 0.07 \\
 log(Ne$^{2+}$/O$^{2+}$)    & -- & -- & -0.95 $\pm$ 0.05 & -- \\
 He$^{+}$/H$^{+}$           &0.052 $\pm$ 0.012 & -- & 0.077 $\pm$ 0.005 & 0.077 $\pm$ 0.005 \\
 ICF(He)                    & 1.15 & -- & 1.19 & 1.20 \\
 He/H                       & 0.060 $\pm$ 0.014 & -- & 0.092 $\pm$ 0.006 & 0.093 $\pm$ 0.006 \\  

\hline
 & & NGC~925 and NGC 1637 & & \\
\hline
\hline 
Parameter  & CDT1 & CDT2 & CDT3 & CDT4 & CDT1 (NGC 1637)\\
            &     &     &       &     & \\
 n$_{e}$    & $\leq$ 40 & $\leq$ 40 & $\leq$ 40 & $\leq$ 40 & 100 \\
$<$log U$>$ & -3.00 $\pm$ 0.30 & -3.26 $\pm$ 0.20 & -3.30 $\pm$ 0.20 & -3.00 $\pm$ 0.25 & -3.30 $\pm$ 0.20 \\
$<t>_{adop}$  & 0.86 & 0.76 & 0.87 & 0.93 & 0.40 \\ 
 12 + log(O$^{2+}$/H$^{+}$) & 7.73 & 7.97 & 7.75 & 7.52 & 8.57 \\
 12 + log(O$^{+}$/H$^{+}$)  & 8.44 & 8.62 & 8.41 & 8.35 & 8.94 \\
 12 + log(O/H)              & 8.52 & 8.71 & 8.50 & 8.41 & 9.10 \\
 12 + log(S$^{2+}$/H$^{+}$) & 6.32 & 6.55 & -- & 6.28 & 6.80 \\
 12 + log(S$^{+}$/H$^{+}$)  & 6.45 & 6.67 & 6.51 & 6.27 & 6.86 \\
 12 + log(S/H)              & 6.69 & 6.91 & --   & 6.58 & 7.13 \\
 log(N/O)                   &-1.07 & --   & --   &-1.01 & -0.88 \\
 log(S/O)                   &-1.83 &-1.80 &  --  &-1.83 & -1.97 \\
\end{tabular}
}
\end{minipage}
\end{table*}
\newpage
                                                         %

\begin{table*}
\setcounter{table}{7}
 \begin{minipage}{150mm}
 \caption{ WR feature intensities and equivalent widths in the observed H{\sevensize II} regions.}
 \footnotesize{
 \begin{tabular}{lcccc}
\hline
\hline
Region & L(WR)/H$\beta$ & EW(WR)({\AA} ) & L(He{\sevensize II})/H$\beta$ & EW(He{\sevensize II})({\AA} ) \\
\hline

Region H13 (NGC 628)   & 0.08 &  8.9 & 0.04  & 4.8 \\ 
Region CDT1 (NGC 1232) & 0.04 &  2.0 & 0.015 & 0.8 \\
Region CDT3 (NGC 1232) & 0.03 &  5.9 & 0.015 & 2.7  \\
Region CDT4 (NGC 1232) & 0.08 & 10   & 0.03  & 3.9 \\
\hline
 \end{tabular}
}
 \end{minipage}
\end{table*}
\newpage
%

                                                                             
\begin{table*}
\setcounter{table}{8}
\begin{minipage}{150mm}
  \caption{Mihalas Single Star Photoionisation Models for the observed H{\sevensize II} regions}
 \scriptsize{
  \begin{tabular}{lcccccccccccc}
  \hline
  \hline
          &  &  &  &  &  NGC 628 &  &  &  &  &  &  &  \\
  \hline
   Region & \multicolumn{2}{c}{H13} & \multicolumn{2}{c}{H3} & \multicolumn{2}{c}{H4} & \multicolumn{2}{c}{H5} &  \multicolumn{2}{c}{} &  \multicolumn{2}{c}{} \\
          & Model & Observed & Model & Observed & Model & Observed & Model & Observed & & & & \\
   Parameter &  &  &  &  &  & & & & & & & \\
  \hline
  \hline
 $<$log U$>$ & -2.70 & -2.78 $\pm$ 0.10 & -2.90 & -2.92 $\pm$ 0.15 & -2.93 & -2.95 $\pm$ 0.15 & -2.97 & -2.97 $\pm$ 0.15 & & & & \\
 n$_{e}$(cm$^{-3}$) & 100 & 80 & 10 & $\leq$ 40 & 10 & $\leq$ 40 & 10 & $\leq$ 40 & & & & \\
 T$_{eff}$(K) & 35,000 & -- & 36,000 & -- & 34,700 & -- & 34,800 & -- & & & & \\
 12 + log(O/H) & 8.24 & 8.24 $\pm$ 0.08 & 8.23 & 8.23 $\pm$ 0.15 & 8.35 & 8.31 $\pm$ 0.15 & 8.35 & 8.34 $\pm$ 0.15 & & & & \\
 12 + log(S/H) & 6.53 & 6.41 $\pm$ 0.04 & 6.52 & 6.38 $\pm$ 0.05 & 6.64 & 6.75 $\pm$ 0.04 & 6.64 & 6.67 $\pm$ 0.05 & & & & \\
 log(N/O) & -1.06 & -1.08 $\pm$ 0.04 & -1.13 & -1.14 $\pm$ 0.10 & -1.15 & -0.94 $\pm$ 0.10 & -1.15 & -1.07 $\pm$ 0.10 & & & & \\
 log(S/O) & -1.71 & -1.83 $\pm$ 0.04 & -1.71 & -1.85 $\pm$ 0.11 & -1.71 & -1.56 $\pm$ 0.11 & -1.71 & -1.67 $\pm$ 0.10 & & & & \\ 
 3727 [OII] & 3519 & 2960 $\pm$ 110 & -- & -- & -- & -- & -- & -- & & & & \\
 5007[OIII] & 1523 & 1547 $\pm$ 10 & 1528 & 1596 $\pm$ 12 & 489 & 488 $\pm$ 8 & 540 & 519 $\pm$ 8 & & & &\\
{\nobreakspace}6584[NII] & 523 & 496 $\pm$ 8 & 477 & 470 $\pm$ 20 & 642 & 603 $\pm$ 40 & 642 & 603 $\pm$ 43 & & & & \\
 6716[SII] & 171 & 204 $\pm$ 5 & 246 & 270 $\pm$ 13 & 305 & 309 $\pm$ 20 & 318 & 333 $\pm$ 24 & & & & \\
 9069[SIII] & 220 & 168 $\pm$ 10 & 213 & 136 $\pm$ 10 & 216 & 147 $\pm$ 12 & 215 & 149 $\pm$ 10 & & & &\\
 4072[SII] & 24 & 29 $\pm$ 3 & -- & -- & -- & -- & -- & -- \\ 
 4363[OIII] & 9 & 10 $\pm$ 2 & 10 & 16 $\pm$ 2 & -- & -- & -- & -- & & & & \\
 5755[NII] & 9 & 5 $\pm$ 1 & -- & -- & -- & -- & -- & -- \\ 
 6312[SIII] & 12 & 12 $\pm$ 1 & 12 & 11 $\pm$ 1 & 10 & 5 $\pm$ 1 & 10 & 7 $\pm$ 2 & & & & \\
 t(O$^{2+}$) & 0.97 & 0.98 $\pm$ 0.05 & 0.99 & 1.17 $\pm$ 0.10 & 0.89 & 0.69 & 0.90 & 0.79 & & & & \\
 t(S$^{2+}$) & 1.00 & 1.02 $\pm$ 0.03 & 1.02 & 1.03 $\pm$ 0.05 & 0.92 & 0.74 $\pm$ 0.04 & 0.93 & 0.82 $\pm$ 0.06 & & & & \\
 t(S$^{+}$) & 0.99 & 0.99 $\pm$ 0.06 & -- & -- & -- & -- & -- & -- & & & & \\
 t(N$^{+}$) & 1.03 & 0.90 $\pm$ 0.06 & -- & -- & -- & -- & -- & -- & & & & \\
  \hline
  \hline
          &  &  &  &  &  NGC 1232 &  &  &  &  &  &  &  \\
  \hline
   Region & \multicolumn{2}{c}{CDT1} & \multicolumn{2}{c}{CDT2} & \multicolumn{2}{c}{CDT3} & \multicolumn{2}{c}{CDT4} &  \multicolumn{2}{c}{} &  \multicolumn{2}{c}{} \\
          & Model & Observed & Model & Observed & Model & Observed & Model & Observed & & & & \\
   Parameter &  &  &  &  &  & & & & & & & \\
  \hline
  \hline
 $<$log U$>$ & -2.85 & -2.95 $\pm$ 0.20 & -3.03 & -2.95 $\pm$ 0.15 & -2.80 & -2.72 $\pm$ 0.10 & -2.65 & -2.72 $\pm$ 0.10 & & & & \\
 n$_{e}$(cm$^{-3}$) & 130 & 130 & 10 & $\leq$ 40 & 230 & 223 & 100 & 118 & & & & \\
 T$_{eff}$(K) & 34,900 & -- & 36,900 & -- & 34,900 & -- & 35,000 & -- & & & & \\
 12 + log(O/H) & 8.97 & 8.95 $\pm$ 0.20 & 8.55 & 8.61 $\pm$ 0.15 & 8.60 & 8.56 $\pm$ 0.14 & 8.49 & 8.37 $\pm$ 0.12 & & & & \\
 12 + log(S/H) & 7.26 & 7.14 $\pm$ 0.10 & 6.89 & 6.80 & 6.96 & 6.86 $\pm$ 0.07 & 6.89 & 6.75 $\pm$ 0.05 & & & & \\
 log(N/O) & -0.87 & -0.81 $\pm$ 0.08 & -1.10 & -- & -1.00 & -0.96 $\pm$ 0.04 & -0.94 & -0.91 $\pm$ 0.06 & & & & \\
 log(S/O) & -1.71 & -1.81 $\pm$ 0.10 & -1.66 & -- & -1.64 & -1.70 $\pm$ 0.07 & -1.60 & -1.62 $\pm$ 0.07 & & & & \\ 
 3727 [OII] & 1782 & 1490 $\pm$ 70 & 4105 & 4180 $\pm$ 220 & 3365 & 3180 $\pm$ 150 & 2925 & 2530 $\pm$ 120 & & & & \\
 5007[OIII] & 275 & 229 $\pm$ 4 & 1347 & 1350 $\pm$ 20 & 830 & 842 $\pm$ 13 & 1259 & 1162 $\pm$ 8 & & & &\\
{\nobreakspace}6584[NII] & 1190 & 1040 $\pm$ 21 & 729 & 717 $\pm$ 30 & 921 & 892 $\pm$ 30 & 760 & 758 $\pm$ 40 & & & & \\
 6716[SII] & 326 & 357 $\pm$ 8 & 470 & 537 $\pm$ 15 & 317 & 333 $\pm$ 10 & 261 & 307 $\pm$ 20 & & & & \\
 9069[SIII] & 299 & 197 $\pm$ 13 & 320 & 191 $\pm$ 15 & 348 & 229 $\pm$ 10 & 342 & 249 $\pm$ 10 & & & &\\
 4072[SII] & -- & -- & -- & -- & 41 & 48 $\pm$ 6 & -- & -- & & & & \\ 
 5755[NII] & 5 & 4 $\pm$ 1 & -- & -- & 9 & 8 $\pm$ 2 & 9 & 8 $\pm$ 1 & & & & \\ 
 6312[SIII] & 4 & 2 $\pm$ 1 & -- & -- & 11 & 8 $\pm$ 2 & 13 & 13 $\pm$ 2 & & & & \\
 t(O$^{2+}$) & 0.55 & 0.45 & 0.81 & -- & 0.76 & 0.69 & 0.79 & 0.84 & & & & \\
 t(S$^{2+}$) & 0.59 & 0.54 $\pm$ 0.05 & 0.85 & -- & 0.80 & 0.74 $\pm$ 0.05 & 0.83 & 0.87 $\pm$ 0.04 & & & & \\
 t(S$^{+}$) & -- & -- & -- & -- & 0.83 & 0.90 $\pm$ 0.06 & -- & -- & & & & \\
 t($N^{+}$) & 0.62 & 0.67 $\pm$ 0.06 & -- & -- & 0.83 & 0.86 $\pm$ 0.06 & 0.88 & 0.90 $\pm$ 0.06 & & & &\\
  \hline
  \hline
          &  &  &  &  &  NGC 925 and NGC 1637 &  &  &  &  &  &  &  \\
  \hline
   Region & \multicolumn{2}{c}{CDT1} & \multicolumn{2}{c}{CDT4} & \multicolumn{2}{c}{CDT1} & \multicolumn{2}{c}{} &  \multicolumn{2}{c}{} &  \multicolumn{2}{c}{} \\
          & Model & Observed & Model & Observed & Model & Observed & & & & & & \\
   Parameter &  &  &  &  &  & & & & & & & \\
  \hline
  \hline
 $<$log U$>$ & -3.05 & -3.00 $\pm$ 0.30 & -3.05 & -3.00 $\pm$ 0.25 & -3.10 & -3.30 $\pm$ 0.20 & & & & & & \\
 n$_{e}$(cm$^{-3}$) & 10 & $\leq$ 40 & 10 & $\leq$ 40 & 100 & 100 & & & & & & \\
 T$_{eff}$(K) & 36,500 & -- & 36,000 & -- & 35,000 & -- & & & & & & \\
 12 + log(O/H) & 8.72 & 8.52 $\pm$ 0.20 & 8.62 & 8.41 $\pm$ 0.20 & 9.17 & 9.10 & & & & & & \\
 12 + log(S/H) & 7.01 & 6.69 & 6.91 & 6.58 & 7.37 & 7.13 & & & & & & \\
 log(N/O) & -0.97 & -1.07 & -0.97 & -1.01 & -0.87 & -0.88 & & & & & & \\
 log(S/O) & -1.71 & -1.83 & -1.71 & -1.83 & -1.80 & -1.97 & & & & & & \\ 
 3727 [OII] & 3414 & 2880 $\pm$ 110 & 3710 & 2820 $\pm$ 120 & 1169 & 1170 $\pm$ 60 & & & & & & \\
 5007[OIII] & 819 & 849 $\pm$ 14 & 784 & 737 $\pm$ 10 & 89 & 84 $\pm$ 10 & & & & & &\\
{\nobreakspace}6584[NII] & 1075 & 568 $\pm$ 28 & 1067 & 593 $\pm$ 18 & 1180 & 1037 $\pm$ 60 & & & & & & \\
 6716[SII] & 469 & 577 $\pm$ 30 & 435 & 437 $\pm$ 15 & 349 & 335 $\pm$ 20 & & & & & & \\
 9069[SIII] & 296 & 125 $\pm$ 9 & 276 & 160 $\pm$ 10 & 201 & 99 $\pm$ 9 & & & & & &\\
 t$_{adopt}$ & 0.70 & 0.86 & 0.75 & 0.93 & 0.46 & 0.40 & & & & & & \\
\hline
\end{tabular}
}
\end{minipage}
\end{table*}

\end{document}

#################################################################

\begin{figure*}
\setcounter{figure}{2}
\begin{minipage}{180mm}
\psfig{figure=h4.epsi,height=10cm,width=17cm,clip=}
\vspace{12pt}
\psfig{figure=h5.epsi,height=10cm,width=17cm,clip=}
\vspace{12pt}
\caption{Merged spectra for the regions H4 and H5 in NGC 628}
\end{minipage}
\end{figure*}

\begin{figure*}
\setcounter{figure}{4}
\begin{minipage}{180mm}
\psfig{figure=CDT3.epsi,height=10cm,width=17cm,clip=}
\vspace{12pt}
\psfig{figure=CDT4.epsi,height=10cm,width=17cm,clip=}   
\caption{Merged spectra for the regions CDT3 and CDT4 in NGC 1232}
\end{minipage}
\end{figure*} 

\begin{figure*}
\setcounter{figure}{5}
\begin{minipage}{180mm}
\psfig{figure=925_1.epsi,height=10cm,width=17cm,clip=}
\vspace{12pt}
\psfig{figure=925_2.epsi,height=10cm,width=17cm,clip=}   
\caption{Merged spectra for the regions CDT1 and CDT2 in NGC 925. 
The absorptions in the Balmer lines and the low signal-to-noise 
ratio is apparent in both regions}
\end{minipage}
\end{figure*} 

\begin{figure*}
\setcounter{figure}{6}
\begin{minipage}{180mm}
\psfig{figure=925_4.epsi,height=10cm,width=17cm,clip=}
\vspace{12pt}
\psfig{figure=1637.epsi,height=10cm,width=17cm,clip=}   
\caption{Merged spectra for regions CDT4 in NGC 925 and CDT1 in NGC 1637. The latter shows a weak [O{\sevensize III}] line intensity}
\end{minipage}
\end{figure*} 
######################################

Another interesting result concerns the observed regions in NGC 925.
These regions, near the nucleus of the galaxy, present several problems.
Despite the observational errors, both the S$_{23}$ abundance calibration and
the discrimination method showed by D00, tend to confirm these regions to have 
an undersolar oxygen content (12 + log(O/H) between 8.5 and 8.7). Certainly,
these values point to a flatter abundance gradient across the disc of this galaxy.\\

The H$\alpha$ fluxes for regions H13, H3, and H5, uncorrected for reddening, are 669, 95 and 62 $\times$ 10$^{-16}$ erg s$^{-1}$ cm$^{-2}$ respectively. 
These fluxes are to be compared with 2089, 372 and 145 $\times$ 10$^{-16}$ erg s$^{-1}$ 
cm$^{-2}$ measured by Kennicutt \& Hodge (1980). H$\alpha$ photographs (Hodge 1975) reveal these regions to be rather extended, and considering that our observations have been obtained
through a 1\farcs03 width slit, hence these lower limits can be considered 
satisfactory. For the rest of the observed regions, no data have been published at the moment to compare with.\\ For each of our observed regions we have 
calculated the H$\alpha$ luminosity, 
the number of hydrogen ionising photons from the reddening corrected H$\alpha$ flux, the filling factor, the H$\alpha$ angular effective diameters, the 
mass of ionised hydrogen and the mass content in the ionising clusters. 
These quantities have been derived according to the expressions given in D00 
and 
are tabulated in Table 17. Regions CDT1, CDT3 and CDT4 in NGC 1232 have 
H$\alpha$ luminosities greater than 10$^{39}$ erg s$^{-1}$ and 
can be classified as supergiant H{\sevensize II} regions as defined by Kennicutt (1983). 
The rest of the regions have H$\alpha$ luminosities typical of H{\sevensize II} regions 
in early spiral galaxies, although all of them are greater than 10$^{37}$ 
erg s$^{-1}$ (Q(H) $>$ 10$^{49}$ photons s$^{-1}$) requiring more than a single
star for their ionisation (Panagia 1973).


Filling factors for each observed region can be determined from the reddening 
corrected  H$\alpha$ flux, F(H$\alpha$), and the derived ionisation parameter, 
U. We have used a value of $\alpha _B(H^o,T)$ = 2.59 $\times$ 10$^{-13}$ cm$^{3}$ s$^{-1}$, corresponding to T= 10000 K and n$_e$= 100 cm$^{-3}$ 
(Osterbrock 1989). For the regions for which only upper limits to the electron density could be derived, a value of n$_e$=10 cm$^{-3}$ has been assumed. 
These values are very similar to those derived in the observed H{\sevensize II} regions 
of NGC 4258 (see D00).
The gas seems to be uniformly distributed throughout these regions. 
The exceptions are regions CDT2 and CDT3 with values around 0.03.


Diameters containing half the H$\alpha$ emission can be obtained from 
the ionization parameter, the reddening corrected H$\alpha$ flux and the derived 
electron density. This angular diameter does not depend on the assumed 
distance to the galaxy. For all the regions, we obtain angular effective diameters between 0.7 and 2.6 arcsec.


The corresponding masses of ionized hydrogen range from 1.12 $\times$ 10$^{4}$ M$_{\odot}$ for region CDT2 in NGC 925 to 9 $\times$ 10$^{4}$ M$_{\odot}$ for 
region CDT4 in NGC 1232.


For our observed regions, all values for Q(H) are greater than 10$^{49}$ photon s$^{-1}$, except for regions CDT1, CDT3 and CDT4 in NGC 1232 for which 
values of the order of 10$^{51}$ photon s$^{-1}$ are found. Hence, in the absence of dust, a lower limit for the mass of the ionising clusters can be 
estimated by means of the H$\beta$ measured equivalent width and the H$\alpha$ luminosity for each region. The estimated ionising cluster masses range from 1.7 $\times$ 10$^{3}$ M$_{\odot}$ for region H4 in NGC 628 to 1.80 $\times$ 10$^{5}$ M$_{\odot}$ for region CDT1 in NGC 1232. It can be inferred from 
these results that all the regions are ionised by small ionising clusters, 
except the three supergiant H{\sevensize II} regions in NGC 1232. Regions H3, H4 and H5 shows a ten to one ratio between their ionised hydrogen mass and the mass 
of the ionising stellar clusters. It can be inferred that these regions are heated by very young ionising clusters. The observed EW(H$\beta$) of 230{\AA} 
supports this claim. 


\begin{figure}
\setcounter{figure}{9}
 \centering
 \psfig{figure=sedh13.epsi,height=8.4cm,width=8.4cm,clip=}
 \vspace{12pt}
 \psfig{figure=sedCDT1.epsi,height=8.4cm,width=8.4cm,clip=}
 \caption{Spectral energy distributions of different ionising clusters for 
regions H13 in NGC 628 (top) and region CDT1 in NGC 1232 (bottom). ST99 stands for 'STARBURST99' (Leitherer at al. 1999)}
\end{figure}


Once the physical conditions have been analysed, the stellar populations 
responsible of the ionisation mechanism are studied in detail. The 
hardening of the ionising radiation due to the presence of Wolf-Rayet stars in 
region H13 has been previously analysed in detail. The presence of 
these stars provides important clues to constrain the age of this 
ionising population. The observed blue and red WR 'bump' luminosities 
relative to H$\beta$ are 0.08 and 0.03. According to the models by Schaerer \& Vacca (
1998), at metallicity Z = 0.004 (0.2 solar), maximum values are 
found for an age of 4Myr (0.06 and 0.03 for the blue and red 'bumps' respectively). 
Moreover, at the age of 4Myr, the blue and red 'bumps' equivalent 
widths are at their maxima, with values around 8.5 {\AA} and 8{\AA} respectively. 
These values are in excellent agreement with the observed ones, 8.9 {\AA} for
the blue bump and 6.1 {\AA} for the red one. As regards as the  individual
lines, the observed values of L(HeII/H$\beta$) and EW(HeII) are 0.04 and 4.8
{\AA} respectively. Again, the agreement can be considered  satisfactory at an
age of 4Myr (0.04 and 4 {\AA} respectively). Finally, models predict a
H$\beta$ equivalent width of 150{\AA} at the age of 4Myr,  to be compared with
the observed value of 140{\AA}. Taking into account  the global and individual
properties of both 'bumps', the agreement  between the models and the derived
quantities is excellent.  In order to predict both the  emission line spectrum
of region H13 and the observed properties  of the WR stars, we have computed
the emission line spectrum  of a single ionising  cluster from the
evolutionary models by Leitherer et al. (1999, STARBURST99)  under the
assumptions of a standard Salpeter initial mass function for the  cluster and
the same metallicity for both gas and ionising stars. Full consistency, within
the errors, is found for a single population of 4.2-4.3Myr.  On the other
hand, single star photoionization models (Mihalas 35,000K and CoStar 36,300K)
also fit both the observed emission line spectrum and ionic abundances, but
this low estimation of the effective temperature  in this region breaks down
the possible  anticorrelation between the effective temperature of the
ionising radiation and metallicity, discussed in D00. However, as it was 
addressed earlier in this section, Wolf Rayet stars in region H13 could be
emitting enough high energy photons, between 35 and 41 eV, to ionize O$^{+}$. 
WR stars are supposed to contribute significantly to the spectral energy
distribution at energies higher than 40 eV, but this contribution is not 
observed in this region (i.e., photoionization of Ne$^{+}$). Hence, this
qualitative interpretation could explain the low estimation of the effective 
temperature from photoionisation models. On the other hand,  though models fit
O$^{2+}$ ionic abundance, a not  observed enhancement in the
Ne$^{2+}$/O$^{2+}$ ratio through the  increase of Ne$^{2+}$ is predicted too.
Spectral energy distributions  for the calculated models in Figure 10(a) show
a non-negligible contribution  of high energy photons between 41 and {$\sim$}
55 eV responsible of that increase.\\

From the five observed H{\sevensize II} regions in NGC 1232, four of them present weak 
blue WR 'bumps' (CDT1, CDT3, CDT4 and CDT5). It must be paid special 
attention onto the high metallicity H{\sevensize II} region CDT1. The observed blue 
WR 'bump' luminosity relative to H$\beta$ is 0.04, extremely low for a high 
metallicity H{\sevensize II} region. A very low value is also found for the blue 'bump' equivalent 
width, near 2 {\AA}. As for the stellar HeII luminosity and 
equivalent width, values of 0.015 and 0.8 {\AA} are found. According to models by 
Schaerer \& Vacca (1998), these low values, together with the observed
H$\beta$ equivalent width of 48 {\AA}, points to an evolved ionising
population  with an age around 7Myr. Evolutionary models with an age of 6-7Myr
predict the emission line spectrum for the derived metallicity, but
underpredict the  observed H$\beta$ equivalent width by a factor of 2. Models
with an age around 5.2Myr,  would fit both the emission line spectrum and the
observed H$\beta$  equivalent width at a higher metallicity (Z/Z$_{\odot}$=2),
but the  predicted properties of the WR bump for this metallicity would be far
overestimated. A possible explanation to this misfit would be the presence of
a  younger stellar ionising population. Figure 10(b) shows the spectral energy
distributions for the calculated models. Despite both the higher  metallicity
derived for this region and the weak WR properties in  comparison to H13, the
shape of the spectral distributions are analogous and this resemblance is
apparent through the same low effective temperatures  derived for both
regions.\\ The same scenario is depicted when this analysis is made onto
regions CDT3 and CDT4. Region CDT3 shows a WR 'bump' luminosity  relative to
H$\beta$ of 0.03. Taking into account the derived metallicity for this region
(12 + logO/H = 8.56 or Z = 0.008), SV98 predicts this value for an H$\beta$
equivalent width of {$\sim$} 200{\AA}, which is in excellent  agreement with
our value of 189{\AA}. At the given metallicity, this equivalent width is
predicted for an instantaneous burst between 3 - 3.5 Myr.  Evenmore, the
predicted blue 'bump' equivalent width at this age range from 4 to 10 {\AA} in
good agreement with the observed one (5.9 {\AA}). Consistency is found also
for the weak red 'bump', which observed values for both the  luminosity and
equivalent width are 0.007 and 2.0{\AA} respectively in  comparison with the
predicted ones ({$\leq$}0.02 and {$\leq$}3 {\AA}). As regards as the HeII line
luminosity and equivalent width, our observed values are 0.015 and 2.7{\AA} in
excellent agreement, again, with the predicted ones ({$\leq$}0.02 and $\sim$3
{\AA}). We have tried to fit both the observed emission  line spectrum and the
derived abundances with several Starburst99 evolutionary  models with ages
ranging from 3 to 4 Myr, and in all of them, though predicted H$\beta$
equivalent widths are consistent with our derived value (250 {\AA} to
150{\AA}), the predicted emission line spectrum results too hard, {\it e.g.}
the  predicted [O{\sevensize III}]{$\lambda$} 5007 {\AA} line intensity relative to
H$\beta$ is,  at least, 4 times higher than the observed value (0.84;
I(H$\beta$ = 1).  In the case of region CDT4, the blue WR 'bump' luminosity
relative to H$\beta$ is 0.08. For the derived metallicity (12 + logO/H = 8.37
or Z = 0.004), this value ($\sim$ 0.06) is predicted to be found for both an
instantaneous burst of 4 Myr and an H$\beta$ equivalent width of {$\sim$}
150{\AA}. Hence, this value is  consistent with the observed one of 138 {\AA}.
The observed blue WR 'bump' equivalent width is 10 {\AA} in good agreement
with the predicted value of 8{\AA} at this age. HeII line intensity and
equivalent width are again fully consistent with the predicted values by SV98
(0.03 versus {$\sim$}0.03 and 3.9{\AA} versus {$\sim$}2{\AA}, respectively).
Again, Starburst99 evolutionary  models with ages ranging from 3.5 to 4.5 Myr,
result too hard to explain the observed emission line spectrum (the predicted
[O{\sevensize III}]{$\lambda$} 5007 {\AA}  line mean intensity relative to H$\beta$ is 3
times higher than the  observed value (1.16; I(H$\beta$ = 1). Both regions
CDT3 and CDT4 can be fit adequately with Mihalas and CoStar single-star
ionization models. These models predict a mean effective temperature for the
ionising clusters of 35000 K (Mihalas) and 36000 K (CoStar) in both regions.\\
 Hence, the four observed supergiant H{\sevensize II} regions show relatively prominent
Wolf-Rayet features with a relative scatter in both the  WR 'bump'
luminosities and equivalent widths. Their derived metallicities range from 
0.2 solar (H13) to 1.25 solar metallicity (CDT1), and the instantaneous bursts
that could explain the observed properties show ages from 3 Myr to 6 Myr.
Despite this observed scatter in both gas and stellar population properties, a
 surprising constant behaviour in the effective temperature of the ionising
radiation  is apparent from our results. Even more surprising is the low
estimation of the effective temperature with a mean value of 35000 K from
Mihalas models or 36000 K  from CoStar models. These results confirm the
derived low effective temperature for the supergiant H{\sevensize II} region 74C in NGC
4258, which also shows a very prominent feature due to Wolf-Rayet stars (see
D00).\\ At this point, the observation of WR features in five supergiant H{\sevensize II}
regions (74C in  NGC 4258, H13 in NGC 628, CDT1, CDT3 and CDT4 in NGC 1232)
provide several clues to understand both the ionization structure of the
nebula and the properties of this  ionising population. However, several
crucial questions arise from this analysis.  The inferred soft hardening of
the ionising radiation in region H13 is the first direct evidence of such
process at the relatively low metallicity regime.  Then it could be expected
this process to be enhanced at higher metallicities, but this scenario is in
contradiction with the low effective temperatures derived in these regions
from single-star photoionization models. A possible explanation is a decrease
in the  predicted number of high energy photons ({$\geq$} 40 eV) emitted by
these stars due to a lower mass loss rate through a higher opacity in the
outer envelopes of these stars. On the other hand, both WR population models
and evolutionary synthesis models fully predict the emission line spectrum,
both the ionic and total heavy element abundances and the  observed WR
properties in region H13. But for regions CDT3, CDT4 and 74C in NGC  4258 (see
D00), while SV98 consistently predict the observed WR properties, evolutionary
models result too hard to explain the observed emission line spectrum. Several
solutions can be suggested such as the inadequacy of the stellar evolutionary
tracks (Luridiana,  Peimbert \& Leitherer 1999), a non-negligible continuous
star formation episode through  few million years instead of a single
instantaneous burst (Garnett ---) or the everlasting temperature fluctuations
within these regions that would underestimate the true abundances. This latter
point is not straightforward from our results. Clearly, an increase in the 
O$^{2+}$ ionic abundance imply a decrease in the mean electron temperature of
the O$^{2+}$ zone that would compensate the predicted steepening of the
electron temperature with radius.  Finally, region CDT1 can be consistently
fit with a lower H$\beta$ equivalent width. In this  case, it could be
possible a composite population to reproduce the emission line spectrum.  
intercorrelation between the ionization 
Wolf-Rayet stellar population is still an open

\section{Summary and conclusions} 

We have analyzed fourteen H{\sevensize II} regions in the galaxies NGC 628, 
NGC 1232, NGC 925 and NGC 1637 using spectrophotometric observations 
between 3800 and 9650 {\AA}. For seven of these regions (H13, H3, H4 and H5 in
 NGC 628 and CDT1, CDT3 and CDT4 in NGC 1232) it has been possible to 
measure mean ion-weighted  temperatures from different 
auroral forbidden lines, 
which allows the derivation of accurate abundances following standard 
methods. For the observed regions in NGC 925, an empirical calibration based on the 
sulphur emission lines has been used to determine a mean oxygen content. 
Region CDT1 in NGC 1637 has been modelled in detail to derive the oxygen 
abundance due to its high metal content.
The derived metallicities range from 0.2 to 0.5 Z$_{\odot}$, except 
region CDT1 in NGC 1232 and CDT1 in NGC 1637 where mean oxygen 
abundances of 8.95 and 9.10 have been derived respectively. As it was 
addressed in D00, from the observed sample of suspected high metallicity
H{\sevensize II} regions, just two of them can be considered regions with a solar
or oversolar abundance. 
Derived mean ion-weighted temperatures from single ionized 
species are shown to be higher than those from the doubled ionized ones. 
Therefore, the general ionization structure proposed by Stasi{\'n}ska (1990) 
and Garnett (1992) for regions in which electron temperatures are lower 
than 10,000K seem to be justified from our results. Region H13 in
 NGC 628 shows a quasi-isothermal behaviour despite the presence 
of Wolf-Rayet stars. On the other hand, region H3 mimics the emission 
line spectrum and physical conditions of its close companion H13. This fact 
allow us to infer a hardening in the ionising radiation due to the
 presence of WR stars in region H13. This hardening is apparent 
through the O$^{2+}$/H$^{+}$ and Ne$^{2+}$/O$^{2+}$ ionic ratios that are 
0.25 dex higher and 0.20 dex lower than in region H3 respectively.  
Evenmore, Wolf Rayet stars in region H13 are emitting enough high
 energy photons ,between 35 and 41 eV, to ionize O$^{+}$, but the 
expected change in the spectral energy distribution due to the 
emission of photons with energies greater than 41eV is not observed from
the derived low value for the Ne$^{2+}$/O$^{2+}$ ionic ratio.  
Region CDT1 in NGC 1232 shows previous trends found in other high 
metallicity H{\sevensize II} regions (M51, D{\'\i}az et al. (1991)), i.e. both N/O 
and S/O ratios higher and lower than solar respectively, 
the former explained if nitrogen 
comes from a secondary production and the latter unexplained in 
high metal environments. \\ From NLTE Costar single star 
photoionisation models, we have also estimated the functional 
parameters for each region, that is, the ionization parameter and the 
effective temperature of the ionising cluster. 
Most regions show ionization parameters of the order of 10$^{-3}$ and 
effective temperatures lower than 37,000 K, except regions H13 in NGC 628, 
and CDT3, CDT4 in NGC 1232 for which 
higher ionization parameters are found. We have also derived the physical 
properties of the regions and their corresponding ionising clusters: 
filling factor, mass of ionized gas and mass of ionising stars. 
Most of the regions have small ionising clusters with masses in the range 1700 to 
30000 solar masses. The exceptions are the three supergiant H{\sevensize II} 
regions in NGC 1232, CDT1, CDT3 and CDT4 with masses greater than 
100,000 solar masses. These values constitute in fact lower limits since the 
regions are assumed to be ionization bounded and the presence of 
dust has not been taken into account.\\WR features have been 
observed in the four supergiant H{\sevensize II} regions,  H13 (NGC 628) and CDT1, CDT3, CDT4 
in NGC 1232. Another fainter Wolf-Rayet feature has been detected in region CDT5. 
A detailed modelling has been carried out for these regions by using two different sets 
of models: those of SV98 for WR populations and those of 
Leitherer et al. (1999) for 
ionising populations, to try to reproduce simultaneously both the WR features 
and the emission line spectrum. In all cases, the agreement between both the
predicted and observed values for the WR luminosities and equivalent widths
is excellent. Furthermore, full consistency is found for 
region H13 with a cluster {$\sim$} 4.2Myr old. In the case of region CDT1, no 
consistent solution is found. A model 7 Myr old can reproduce both the 
emission line spectrum and the WR properties, but not the observed 
H$\beta$ equivalent width. Conversely, a model 5.2 Myr old reproduce both 
the emission line spectrum and the H$\beta$ equivalent width at twice 
solar metallicity but overestimates by far the observed WR properties. 
On the other hand, Starburst99 evolutionary synthesis
models with ages ranging from 3 to 4.5 Myr, result too hard to explain the
observed emission line spectrum in both regions CDT3 and CDT4.\\ 
Moreover, from single-stars photoionization models, these observed regions with 
both different WR properties and metallicities show the same low estimation of the effective
temperature, around 36,000K (CoStar models) or 35,000K (Mihalas models) hence, from
these results, there is no evidence for a correlation between the effective 
temperature of the ionising radiation and metallicity (D00). 
Though a soft hardening of the ionising radiation is apparent in region H13, 
the expected hardening from theoretical grounds in the spectral energy
distribution might happen in a short 
period of time, hence it would be difficult to observe. Again, the observation
of WR features belonging to a 4 Myr cluster population in giant extragalactic
high metallicity H{\sevensize II} regions (like region CDT1 in NGC 1232) would help to 
clarify the relationship between the hardening of the ionising radiation and the 
ionisation structure in these regions.

\section*{Acknowledgements}

The WHT is operated in the island of La Palma by the Issac Newton Group 
in the Spanish Observatorio del 
Roque de los Muchachos of the Instituto de Astrof{\'\i}sica de Canarias. We 
would like to thank CAT for awarding observing time.

E.T. is grateful to an IBERDROLA Visiting Professorship 
to UAM during which part of this work was completed.
This work has been partially supported by DGICYT project PB-96-052.




%


\newpage

%
%

\begin{table*}
\setcounter{table}{11}
 \begin{minipage}{150mm}
  \caption{Single star and evolutionary models for region H13 in NGC 628}
  \begin{tabular}{@{}lcccc@{}}
   Parameter & Observations & Mihalas 35,000K & CoStar 36,300K & 4.3 Myr Model (ST99) \\
 3727 [OII]  & 2960 $\pm$ 110 & 3500 & 3358 & 3410 \\
 4363 [OIII] & 10 $\pm$  2 & 9 & 11 & 12 \\
 4959 [OIII] & 526 $\pm$  5 & 527 & 577 & 586 \\
 5007 [OIII] & 1547 $\pm$ 10 & 1520 & 1665 & 1692 \\
 5755 [NII]  & 5 $\pm$ 1 & 9 & 9 & 9 \\
 6548 [NII]  & 160 $\pm$  3 & 176 & 171 & 174 \\ 
 6584 [NII]  & 496 $\pm$  8 & 519 & 503 & 513 \\
 4072 [SII]  & 29 $\pm$ 3 & 25 & 29 & 30 \\
 6717 [SII]  & 204 $\pm$  5 & 173 & 203 & 202 \\
 6731 [SII]  & 151 $\pm$  4 & 130 & 152 & 151 \\
 6312 [SIII] & 12 $\pm$  1 & 13 & 12 & 11 \\
 9069 [SIII] & 168 $\pm$ 10 & 222 & 199 & 179 \\
 9532 [SIII] & 374 $\pm$ 15 & 551 & 496 & 445 \\ 
 EW(H$\beta$)({\AA}) & 140 & -- & -- & 123 \\
 log Q(H) & 50.69 & -- & -- & 50.69 \\
 $<$log U$>$ & -- & -2.70 & -2.83 & -2.85 \\
 n$_{e}$ & 80 & 100 & 100 & 100 \\
 t(N$^{+}$)  & 0.90 $\pm$ 0.07 & 1.03 & 1.06 & 1.07 \\
 t(S$^{+}$)  & 0.99 $\pm$ 0.09 & 1.00 & 0.99 & 1.00 \\
 t(O$^{2+}$) & 0.98 $\pm$ 0.06 & 0.97 & 1.01 & 1.03 \\
 t(S$^{2+}$) & 1.02 $\pm$ 0.06 & 1.01 & 0.99 & 1.05 \\
 12 + log(O$^{2+}$/H$^{+}$) & 7.77 $\pm$ 0.09 & 7.71 & 7.68 & 7.65 \\
 12 + log(O$^{+}$/H$^{+}$) & 8.06 $\pm$ 0.19 & 8.03 & 7.94 & 7.92 \\
 12 + log(O/H) & 8.24 $\pm$ 0.15 & 8.23 & 8.18 & 8.16 \\
 12 + log(S$^{2+}$/H$^{+}$) & 6.25 $\pm$ 0.06 & 6.41 & 6.31 & 6.25 \\
 12 + log(S$^{+}$/H$^{+}$) & 5.91 $\pm$ 0.10 & 5.88 & 5.96 & 5.94 \\
 12 + log(S/H) & 6.41 $\pm$ 0.09 & 6.53 & 6.48 & 6.43 \\
 log(S/O) & -1.83 $\pm$ 0.07 & -1.72 & -1.72 & -1.74 \\
 log(N/O) & -1.08 $\pm$ 0.10 & -1.06 & -1.06 & -1.05 \\
 log(Ne$^{2+}$/O$^{2+}$)    &-1.04 $\pm$ 0.04 & -0.76 & -0.65 & -0.70 
\end{tabular}
\end{minipage}
\end{table*}
\newpage
\begin{table*}
\setcounter{table}{12}
 \begin{minipage}{150mm}
  \caption{Single star and evolutionary models for region CDT1 in NGC 1232}
  \begin{tabular}{@{}lcccc@{}}
   Parameter & Observations & CoStar 36,000K & 6.0 Myr Model (ST99) & 5.2 Myr Model (ST99) \\
 3727 [OII]  & 1490 $\pm$ 70 & 1650 & 1649 & 1531 \\
 4959 [OIII] & 77 $\pm$  2 & 84 & 87 & 91 \\
 5007 [OIII] & 229 $\pm$ 4 & 242 & 251 & 265 \\
 5755 [NII]  & 4 $\pm$ 1 & 4 & 4 & 5 \\
 6548 [NII]  & 328 $\pm$ 8 & 356 & 350 & 363 \\ 
 6584 [NII]  & 1040 $\pm$ 21 & 1050 & 1034 & 1072 \\
 6717 [SII]  & 357 $\pm$ 8 & 386 & 374 & 390 \\
 6731 [SII]  & 274 $\pm$ 6 & 294 & 285 & 295 \\
 6312 [SIII] & 2: & 3 & 2 & 2 \\
 9069 [SIII] & 197 $\pm$ 13 & 290 & 184 & 142 \\
 9532 [SIII] & 455 $\pm$ 34 & 719 & 456 & 358 \\ 
 EW(H$\beta$)({\AA}) & 48 & -- & 19 & 45 \\
 log Q(H) & 51.18 & -- & 51.18 & 51.18 \\
 $<$log U$>$ & -- & -2.90 & -2.97 & -2.97 \\
 n$_{e}$ & 130 & 100 & 100 & 100 \\
 t(N$^{+}$)  & 0.67 $\pm$ 0.07 & 0.60 & 0.61 & 0.55\\
 t(S$^{+}$)  & -- & 0.59 & 0.61 & 0.54 \\
 t(O$^{2+}$) & 0.45 & 0.50 & 0.52 & 0.34\\
 t(S$^{2+}$) & 0.54 $\pm$ 0.09 & 0.55 & 0.56 & 0.42 \\
 12 + log(O$^{2+}$/H$^{+}$) & 8.63 $\pm$ 0.50 & 8.14 & 8.10 & 8.66 \\
 12 + log(O$^{+}$/H$^{+}$) & 8.66 $\pm$ 0.33 & 8.94 & 8.88 & 9.11 \\
 12 + log(O/H) & 8.95 $\pm$ 0.40 & 9.05 & 9.00 & 9.35 \\
 12 + log(S$^{2+}$/H$^{+}$) & 6.97 $\pm$ 0.30 & 7.15 & 6.92 & 7.16 \\
 12 + log(S$^{+}$/H$^{+}$) & 6.67 $\pm$ 0.20 & 6.90 & 6.86 & 7.16 \\
 12 + log(S/H) & 7.14 $\pm$ 0.20 & 7.34 & 7.19 & 7.46 \\
 log(S/O) & -1.81 $\pm$ 0.20 & -1.71 & -1.80 & -1.87 \\
 log(N/O) & -0.81 $\pm$ 0.12 & -0.92 & -0.92 & -0.92
\end{tabular}
\end{minipage}
\end{table*}

\newpage

%
%
%

\begin{table*}
\setcounter{table}{13}
 \begin{minipage}{150mm}
  \caption{Single-star photoionization models for regions H3, H4 and H5 in NGC 628}
  \begin{tabular}{@{}lccccccccc@{}}
   & & & & Region & & & & & \\
   &  & H3 &  &  & H4 & & & H5 & \\
Parameter    & Observed & & Model & Observed & & Model & Observed & & Model \\
 4363 [OIII] & 16 $\pm$ 2 & & 14 & -- & & 4 & -- & & 4   \\
 4959 [OIII] & 510 $\pm$ 6 & & 563 & 166 $\pm$ 4 & & 159 & 175 $\pm$ 4 & & 190 \\
 5007 [OIII] & 1596 $\pm$ 12 & & 1620 & 488 $\pm$ 8 & & 460 & 519 $\pm$ 8 & & 560 \\
 6548 [NII]  & 162 $\pm$ 6 & & 174 & 205 $\pm$ 15 & & 213 & 205 $\pm$ 17 & & 210 \\ 
 6584 [NII]  & 470 $\pm$ 20 & & 512 & 603 $\pm$ 40 & & 628 & 603 $\pm$ 43 & & 620 \\
 6717 [SII]  & 270 $\pm$ 13 & & 240 & 309 $\pm$ 20 & & 270 & 333 $\pm$ 24 & &305 \\
 6731 [SII]  & 186 $\pm$ 9 & & 167 & 215 $\pm$ 15 & & 182 & 227 $\pm$ 17 & & 201 \\
 6312 [SIII] & 11 $\pm$ 2 & & 12 & 5 $\pm$ 1 & & 10 & 7 $\pm$ 2 & & 10 \\
 9069 [SIII] & 136 $\pm$ 10 & & 176 & 147 $\pm$ 12 & & 237 & 149 $\pm$ 10 & & 241 \\
 9532 [SIII] & 351 $\pm$ 25 & & 436 & 385 $\pm$ 23 & & 588 & 399 $\pm$ 20 & & 606\\ 
 $<$log U$>$ & -- & & -3.00 & -- & & -2.70 & -- & & -2.75\\
 T$_{eff}$ & -- & & 36,600K & -- & & 35,600K & -- & & 35,900K\\
 n$_{e}$ & $\leq$ 40 & & 10 & $\leq$ 40 & & 10 & $\leq$ 40 & & 10 \\ 
 t(O$^{+}$)  & 1.10 & & 1.11 & 0.79 & & 0.90 & 0.85 & & 0.92 \\
 t(O$^{2+}$) & 1.17 $\pm$ 0.09 & & 1.07 & 0.69 & & 0.85 & 0.79 & & 0.86 \\
 t(S$^{2+}$) & 1.03 $\pm$ 0.10 & & 1.09 & 0.74 $\pm$ 0.08 & & 0.89 & 0.82 $\pm$ 0.08 & & 0.91 \\
 12 + log(O$^{2+}$/H$^{+}$) & 7.53 $\pm$ 0.08 & & 7.55 & 7.89 $\pm$ 0.30 & & 7.41 & 7.66 $\pm$ 0.23 & & 7.59 \\
 12 + log(O$^{+}$/H$^{+}$) & 7.96 $\pm$ 0.16 & & 7.85 & 8.32 $\pm$ 0.13 & & 8.29 & 8.29 $\pm$ 0.13 & & 8.18 \\
 12 + log(O/H) & 8.10 $\pm$ 0.15 & & 8.10 & 8.46 $\pm$ 0.15 & & 8.35 & 8.38 $\pm$ 0.15 & & 8.32 \\ 
 12 + log(S$^{2+}$/H$^{+}$) & 6.20 $\pm$ 0.10 & & 6.19 & 6.52 $\pm$ 0.14 & & 6.56 & 6.43 $\pm$ 0.12 & & 6.51 \\
 12 + log(S$^{+}$/H$^{+}$) & 5.92 $\pm$ 0.07 & & 5.97 & 6.35 $\pm$ 0.13 & & 6.11 & 6.29 $\pm$ 0.12 & & 6.24 \\
 12 + log(S/H) & 6.38 $\pm$ 0.09 & & 6.40 & 6.75 $\pm$ 0.15 & & 6.69 & 6.67 $\pm$ 0.12 & & 6.70 \\
 log(S/O) & -1.71 & & -1.71 & -1.71 & & -1.67 & -1.71 & & -1.63 \\
 log(N/O) & -0.93 & & -1.03 & -0.93 & & -1.10 & -0.93 & & -1.07
\end{tabular}
\end{minipage}
\end{table*}
\newpage
%

\begin{table*}
\setcounter{table}{14}
 \begin{minipage}{150mm}
  \caption{Single-star photoionization models for regions CDT2, CDT3 and CDT4 in NGC 1232}
  \begin{tabular}{@{}lccccccccc@{}}
   & & & & Region & & & & & \\
   &  & CDT2 &  &  & CDT3 & & & CDT4 & \\
Parameter    & Observed & & Model & Observed & & Model & Observed & & Model \\
 3727 [OII]  & 4180 $\pm$ 220 & & 4280 & 3180 $\pm$ 150 & & 3130 & 2530 $\pm$ 120 & & 2720 \\
 4959 [OIII] & 456 $\pm$ 8 & & 480 & 284 $\pm$ 6 & & 273 & 393 $\pm$ 3 & & 417 \\
 5007 [OIII] & 1350 $\pm$ 20 & & 1386 & 842 $\pm$ 13 & & 789 & 1162 $\pm$ 8 & & 1205 \\
 5755 [NII]  & -- & & & 8 $\pm$ 2 & & 9 & 8 $\pm$ 1 & & 9 \\
 6548 [NII]  & 231 $\pm$ 15 & & 253 & 290 $\pm$ 10 & & 318 & 252 $\pm$ 14 & & 270 \\
 6584 [NII]  & 717 $\pm$ 30 & & 746 & 892 $\pm$ 30 & & 936 & 758 $\pm$ 40 & & 798 \\
 4072 [SII]  & -- & & -- & 48 $\pm$ 6  & & 38 & -- & & -- \\
 6717 [SII]  & 537 $\pm$ 15 & & 515 & 333 $\pm$ 10 & & 297 & 307 $\pm$ 20 & & 275 \\
 6731 [SII]  & 372 $\pm$ 10 & & 362 & 272 $\pm$ 9 & & 250 & 233 $\pm$ 15 & & 209 \\
 6312 [SIII] & -- & & & 8 $\pm$ 2 & & 10 & 13 $\pm$ 2 & & 12 \\
 9069 [SIII] & 191 $\pm$ 15 & & 297 & 229 $\pm$ 10 & & 301 & 249 $\pm$ 10 & & 320 \\
 9532 [SIII] & 535 $\pm$ 40 & & 737 & 614 $\pm$ 30 & & 747 & 665 $\pm$ 20 & & 793 \\
 $<$log U$>$ & -- & & -3.10 & -- & & -2.76 & -- & & -2.68 \\
 T$_{eff}$ & -- & & 38,700K & -- & & 35,800K & -- & & 36,000K \\
 n$_{e}$ & $\leq$ 40 & & 10 & 223 & & 230 & 118 & & 100 \\
 t(S$^{+}$)  & -- & & 0.90 & $\leq$ 0.90 & & 0.84 & -- & & 0.87 \\
 t(N$^{+}$)  & -- & & 0.93 & 0.86 $\pm$ 0.07 & & 0.84 & 0.90 $\pm$ 0.07 & & 0.88 \\
 t(O$^{2+}$) & 0.81 & & 0.84 & 0.69 & & 0.76 & 0.84 & & 0.79 \\
 t(S$^{2+}$) & 0.84 & & 0.88 & 0.74 $\pm$ 0.07 & & 0.80 & 0.87 $\pm$ 0.06 & & 0.83 \\
 12 + log(O$^{2+}$/H$^{+}$) & 8.03 & & 7.82 & 8.13 $\pm$ 0.25 & & 7.81 & 7.90 $\pm$ 0.12 & & 7.92 \\
 12 + log(O$^{+}$/H$^{+}$) & 8.46 & & 8.28 & 8.37 $\pm$ 0.16 & & 8.43 & 8.20 $\pm$ 0.14 & & 8.25 \\
 12 + log(O/H) & 8.61 $\pm$ 0.15 & & 8.50 & 8.56 $\pm$ 0.20 & & 8.55 & 8.37 $\pm$ 0.12 & & 8.43 \\
 12 + log(S$^{2+}$/H$^{+}$) & 6.53 & & 6.59 & 6.72 $\pm$ 0.13 & & 6.75 & 6.60 $\pm$ 0.08 & & 6.74 \\
 12 + log(S$^{+}$/H$^{+}$) & 6.47 & & 6.48 & 6.30 $\pm$ 0.17 & & 6.36 & 6.21 $\pm$ 0.09 & & 6.27 \\
 12 + log(S/H) & 6.80 & & 6.84 & 6.86 $\pm$ 0.13 & & 6.90 & 6.75 $\pm$ 0.08 & & 6.87 \\
 log(S/O) & -1.81 & & -1.66 & -1.70 $\pm$ 0.07 & & -1.66 & -1.62 $\pm$ 0.10 & & -1.60 \\
 log(N/O) & -1.15 & & -1.10 & -0.96 $\pm$ 0.12 & & -0.97 & -0.91 $\pm$ 0.08 & & -0.91 
\end{tabular}
\end{minipage}
\end{table*}
%
%

\begin{table*}
\setcounter{table}{15}
 \begin{minipage}{150mm}
  \caption{Single-star photoionization models for regions CDT1, CDT4 in NGC 925  and CDT1 in NGC 1637}
  \begin{tabular}{@{}lccccccccc@{}}
   & & & & Region & & & & & \\
   &  & CDT1 &  &  & CDT4 & & & CDT1 & \\
Parameter    & Observed & & Model & Observed & & Model & Observed & & Model \\
 3727 [OII]  & 2880 $\pm$ 110 & & 2770 & 2820 $\pm$ 120 & & 2896 & 1170 $\pm$ 60 & & 1390 \\
 4959 [OIII] & 301 $\pm$ 7 & & 290 & 239 $\pm$ 4 & & 257 & 29 $\pm$ 2 & & 29 \\
 5007 [OIII] & 849 $\pm$ 14 & & 837 & 737 $\pm$ 10 & & 741 & 84 $\pm$ 10 & & 86 \\
 6548 [NII]  & 203 $\pm$ 11 & & 202 & 196 $\pm$ 7 & & 219 & 315 $\pm$ 22 & & 359 \\
 6584 [NII]  & 568 $\pm$ 28 & & 595 & 593 $\pm$ 18 & & 648 & 1037 $\pm$ 60 & & 1060 \\
 6717 [SII]  & 577 $\pm$ 30 & & 547 & 437 $\pm$ 15 & & 440 & 335 $\pm$ 20 & & 342 \\
 6731 [SII]  & 391 $\pm$ 20 & & 377 & 298 $\pm$ 11 & & 307 & 249 $\pm$ 15 & & 261 \\
 9069 [SIII] & 125 $\pm$ 9 & & 362 & 160 $\pm$ 10 & & 295 & 99 $\pm$ 9 & & 292\\
 9532 [SIII] & 359 $\pm$ 25 & & 897 & 342 $\pm$ 25 & & 731 & 247 $\pm$ 15 & & 725 \\
 $<$log U$>$ & -- & & -2.95 & -- & & -2.95 & -- & & -2.77 \\
 T$_{eff}$ & -- & & 37,500K & -- & & 36,300K & -- & & 35,000K \\
 n$_{e}$ & $\leq$ 40 & & 10 & $\leq$ 40 & & 10 & 100 & & 100 \\
 t(O$^{2+}$) & 0.86 & & 0.64 & 0.93 & & 0.65 & -- & & 0.48 \\
 12 + log(O$^{2+}$/H$^{+}$) & 7.73 & & 8.11 & 7.52 & & 8.04 & -- & & 7.86 \\
 12 + log(O$^{+}$/H$^{+}$) & 8.44 & & 8.55 & 8.35 & & 8.59 & -- & & 9.05 \\
 12 + log(O/H) & 8.52 & & 8.75 & 8.41 & & 8.75 & -- & & 9.09 \\
 12 + log(S$^{2+}$/H$^{+}$) & 6.32 & & 6.95 & 6.28 & & 6.86 & -- & & 7.22 \\
 12 + log(S$^{+}$/H$^{+}$) & 6.45 & & 6.74 & 6.27 & & 6.63 & -- & & 6.91 \\
 12 + log(S/H) & 6.69 & & 7.16 & 6.58 & & 7.06 & -- & & 7.39 \\
 log(S/O) & -1.83 & & -1.60 & -1.83 & & -1.70 & -- & & -1.71 \\
 log(N/O) & -1.07 & & -1.15 & -1.01 & & -1.15 & -- & & -0.92 
\end{tabular}
\end{minipage}
\end{table*}


\begin{table*}
\setcounter{table}{10}
 \begin{minipage}{150mm}
 \caption{ WR feature intensities and equivalent widths in the observed H{\sevensize II} regions.}
 \begin{tabular}{lcccccc}
Region & L(WR)/H$\beta$ & EW(WR)({\AA} ) & L(HeII)/H$\beta$ & EW(HeII)({\AA} ) & 
12+log(O/H) & T$_{eff}$ (K) \\
Region H13 (NGC 628)   & 0.08 &  8.9 & 0.04 & 4.8 & 8.24 & 36300 \\ 
Region CDT1 (NGC 1232) & 0.04 &  2.0 & 0.015 & 0.8 & 8.95 & 36000 \\
Region CDT3 (NGC 1232) & 0.03 &  5.9 & 0.015 & 2.7 & 8.56 & 35800 \\
Region CDT4 (NGC 1232) & 0.08 &   10 & 0.03 & 3.9 & 8.37 & 36000
 \end{tabular}
 \end{minipage}
 \end{table*}


\newpage

\begin{table*}
\setcounter{table}{16}
 \begin{minipage}{120mm}
 \caption{Physical properties of the observed H{\sevensize II} regions}
 \begin{tabular}{llcccccc}
Galaxy & Region & L(H$\alpha$) & Q(H) & $\epsilon$ & M(H{\sevensize II}) & M$^{\star}$ & $\phi$ \\
       &        & (10$^{38}$ erg s$^{-1}$) & (10$^{49}$ s$^{-1}$) & & (M$_{\odot}$) & (M$_{\odot}$) & (arcsec)\\
NGC 628 & H13   &  6.63   & 48.5     & 0.845  & 17900 & 22912 & 1.6 \\
       & H3    &  1.19   & 8.73     & 0.43   & 25800 &  2680 & 2.6 \\
       & H4    &  0.695  & 5.09     & 0.65   & 15000 &  1720 & 1.6 \\
       & H5    &  0.81   & 5.88     & 0.50   & 17300 &  1920 & 1.9 \\
NGC 925 & CDT1  &  0.91   & 6.67     & 0.41   & 19700 & 15475 & 1.5 \\
       & CDT2  &  0.59   & 4.28     & 0.26   & 12600 & 30355 & 1.5 \\
       & CDT3  &  0.82   & 6.02     & 0.15   & 17800 & 19420 & 2.0 \\
       & CDT4  &  1.27   & 9.27     & 0.25   & 27400 & 11196 & 2.0 \\
NGC 1232 & CDT1 &  20.8   & 152      & 0.02  & 20800 & 180290 & 0.9\\
        & CDT2 &  2.27   & 16.6     & 0.03   & 49000 & 12168  & 1.6 \\
        & CDT3 &  38.5   & 282      & 0.03   & 25200 & 102900 & 0.7 \\
        & CDT4 &  48.2   & 353      & 0.26   & 88300 & 168841 & 1.1 \\
        & CDT5 &  6.30   & 46.2     & --     & 16400 & 43372 & \\
NGC 1637 & CDT1 &  1.77   & 13.0     & 0.10  &  2650 & 10586 & 1.5      
 \end{tabular}
 \end{minipage}
\end{table*}

\newpage
%
%
%

\end{document}